\documentclass{aa}  
\usepackage{graphicx}
\usepackage{natbib}
\usepackage{txfonts}
\usepackage{xcolor}
\usepackage[breaklinks, colorlinks, citecolor=blue]{hyperref}
\usepackage{currvita}
\usepackage{eqnarray,amsmath}

\DeclareMathAlphabet{\mathsc}{OT1}{cmr}{m}{sc}

\newcommand{\Planck}{{\it Planck}}
\newcommand{\HI}{\ensuremath{\mathsc {H\,i}}}
\newcommand{\HII}{\ensuremath{\mathsc {H\,ii}}}
\newcommand{\ArI}{\ensuremath{\mathsc {A{\rm r}\, i}}}
\newcommand{\OI}{\ensuremath{\mathsc {O\, i}}}
\newcommand{\CII}{\ensuremath{\mathsc {C\, ii}}}
\newcommand{\NII}{\ensuremath{\mathsc {N\, ii}}}

\begin{document} 

\title{Associating LOFAR Galactic Faraday structures with the warm neutral medium}

  \author{F. Boulanger \inst{1}\thanks{Corresponding author: francois.boulanger@ens.fr} 
  \and C. Gry \inst{2}
  \and E. B. Jenkins \inst{3}
  \and A. Bracco \inst{1,4}
\and A. Erceg\inst{5} 
\and V. Jeli\'{c}  \inst{5} 
\and L. Turi\'{c}\inst{5}
 } 
   
   \institute{Laboratoire de Physique de l'Ecole Normale Sup\'erieure, ENS, Universit\'e PSL, CNRS, Sorbonne Universit\'e, Universit\'e de Paris, F-75005 Paris, France
   \and
   Aix Marseille Univ, CNRS, CNES, LAM, Marseille, France
   \and
   Department of Astrophysical Sciences, Princeton University, Princeton NJ 08544, USA
   \and
   INAF–Osservatorio Astrofisico di Arcetri, Largo E. Fermi 5, 50125 Firenze, Italy
    \and
   Ru{\dj}er Bo\v{s}kovi\'{c} Institute, Bijeni\v{c}ka cesta 54, 10000 Zagreb, Croatia       
   }

   \date{Received December 14, 2023; accepted April 18, 2024}

\abstract {Faraday tomography observations with the Low Frequency Array (LOFAR) have unveiled a remarkable network of structures in polarized synchrotron emission at high Galactic latitudes. 
The observed correlation between LOFAR structures, dust polarization, and \HI\ emission suggests a connection to the neutral interstellar medium (ISM). We investigated this relationship by estimating the rotation measure (RM) of the warm neutral (partially ionized) medium (WNM) in the local ISM. Our work combines UV spectroscopy from FUSE and dust polarization observations from \Planck\ with LOFAR data. We derived electron column densities from UV absorption spectra toward nine background stars, within the field of published data from the LOFAR two-meter sky survey. The associated RMs were estimated using a local magnetic field model fitted to the dust polarization data of \Planck. A comparison with Faraday spectra at the position of the stars suggests that LOFAR structures delineate a slab of magnetized WNM and synchrotron emission, located ahead of the bulk of the warm ionized medium. This conclusion establishes an astrophysical framework for exploring the link between Faraday structures and the dynamics of the magnetized multiphase ISM. It will be possible to test it on a larger sample of stars when maps from the full northern sky survey of LOFAR become available.  }


%\offprints{\url{francois.boulanger@ens.fr}}
\authorrunning{F. Boulanger et al.}
\titlerunning{LOFAR Faraday structures from the warm neutral medium}

   \maketitle
\section{Introduction}

New telescopes, instruments, and sky surveys in low-frequency radio astronomy \citep{Wayth15,Shimwell17,Wolleben19,Shimwell19,Shimwell22} are poised to greatly expand studies of the diffuse interstellar medium (ISM). Rotation measure (RM) synthesis
emphasizes the information encoded in the observations by distinguishing polarized emission along the line of sight (LoS) based on the Faraday rotation it has undergone \citep{Brentjens&deBruyn:2005,vanEck18}. 
This data analysis technique, which provides data cubes of polarized intensity versus Faraday depth, is referred to as Faraday tomography.  The data cubes reveal an unexpected structural richness in the polarized Galactic synchrotron emission that offers information on the 3D structure of interstellar magnetic fields and electrons  \citep[][]{Iacobelli13,Jelic:2014,Jelic:2015,Lenc16,vanEck:2017,Thomson19,vanEck19, Turic21, Erceg22}.
When interpreting the data, we are faced with our limited understanding of the ionization of the diffuse ISM and, in particular, with the difficulty in identifying the ISM component to which the Faraday structures relate. 

Correlation with ISM tracers provides clues on the nature of Faraday features.  
Comparison of LOFAR observations with Planck maps have revealed a tight correlation between the orientation of structures in Faraday data and the magnetic field component in the plane of the sky traced by dust polarization in a few mid-latitude Galactic fields located in the surroundings of the extragalactic point source 3C~196  \citep{Zaroubi:2015,Jelic:2018,Turic21}. 
LOFAR structures have also been observed to correlate with \HI\ filaments associated with the cold neutral medium \citep{Kalberla16,Kalberla:2017,Jelic:2018} and, more generally, with maps of ISM phases (cold, unstable and warm gas) derived from the analysis of \HI\ spectral data cubes \citep{Bracco20}. 
The observational results are limited because the published data comparisons focus on a few fields close together in the sky\footnote{ The correlation between dust polarization and Faraday rotation should at least depend on the average orientation of the magnetic field, as seen in the study by Bracco et al. (2022) using synthetic data.}. However, they point to a common physical framework in which the LOFAR structures trace the neutral dusty ISM, and not the warm ionized medium (WIM). To test this hypothesis, we present this work in which we estimate the RM associated with the warm neutral medium (WNM) in the local ISM.

The UV spectroscopic data in absorption toward background stars indicate that the ionization fraction of the WNM is about 0.1, in the thick Galactic disk \citep{Spitzer93}, the local ISM  \citep{Jenkins13} and small clouds of warm gas embedded within the Local Bubble\footnote{The low density interstellar region surrounding the Sun to distances of 100 to $300\,$pc \citep{Lallement14,Zucker22}} \citep[LB,][]{Redfield08b}, in particular the local interstellar cloud encompassing the Sun  \citep{Slavin08,Gry17,Bzowski19}). The WNM, rather than the WIM  \citep{Reynolds:1998},
could be the specific ISM component traced by low-frequency Faraday observations at high Galactic latitude, as proposed by \citet{Heiles_Haverkorn12} and \citet{vanEck:2017}, and supported by the numerical simulations presented by \citet{Bracco:2022}.

This work brings together the analysis of the LOFAR Two-meter Sky
Survey \citep[LoTSS,][]{Shimwell22}, the mosaic at high latitudes toward the outer Galaxy analyzed by \citet{Erceg22}, with UV spectroscopy and dust polarization observations. 
We combine electron column densities derived from UV spectroscopic observations toward stars with a model of the local magnetic field fitted on \Planck\ dust 
polarization maps to estimate the RM of the WNM in the local ISM. The comparison with Faraday spectra on the same LoS allows us to assess the WNM contribution to the Faraday rotation of the polarized synchrotron emission measured by LOFAR.   

\begin{figure*}[!h]
\centering
\includegraphics[scale=0.45]{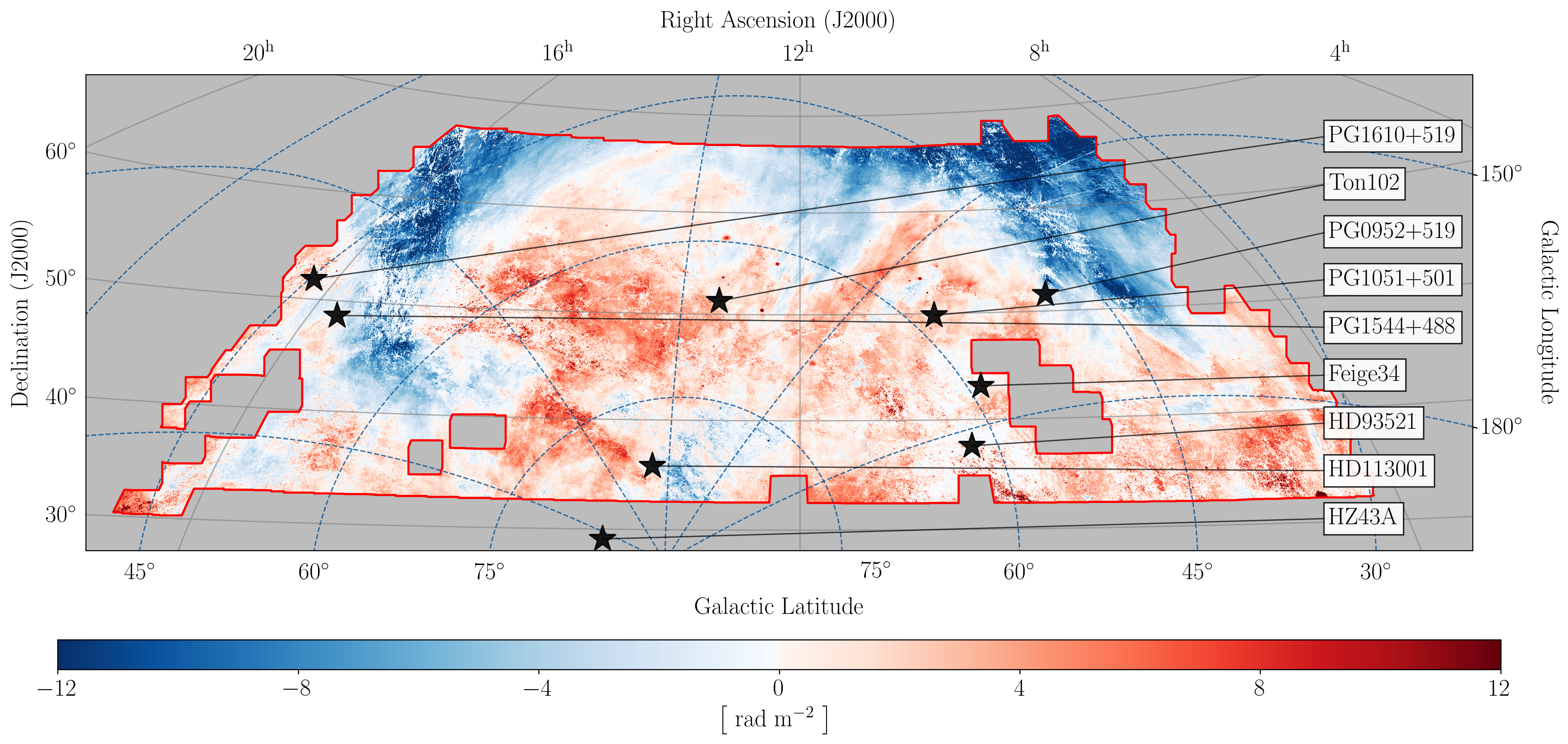}
\caption{Stars in our sample overlaid on a LOFAR image of the LoTSS field from \citet{Erceg22}. The image displays the first moment of the Faraday spectra 
in units of ${\rm rad~m^{-2}}$. 
Celestial and Galactic coordinates are plotted with solid and dashed lines, respectively.  }
\label{fig:LOFAR_map}
\end{figure*}

The paper is organized as follows. We introduce stellar targets and present the UV observational data in Sect.~\ref{sec:stars}. The model of the local magnetic field that we use to estimate the RM from electrons in the WNM is presented in Sect.~\ref{sec:Bfield_model} and Appendix~\ref{App:Bfield}. 
The comparison with LOFAR observations is presented in Sect.~\ref{sec:LOFAR}. 
Our work brings a new perspective to the interpretation of LOFAR Faraday data: the origin of the structures and the contribution from the WIM, which we discuss in
Sect.~\ref{sec:insight}.
The main results are summarized in Sect.~\ref{sec:conclusion}.

\begin{table*}
%\centering
{\small
    \caption{Stellar targets and results of the data fit}
\label{stellar_data}
   \begin{tabular}{lccccccc}
    \hline
    \hline
    \\[-1.0ex]
 Star    & $l $ & $b$ & parallax & d  & log($N{(\OI})$) & b & [\ArI/\OI]\\
        & $^\circ$  & $^\circ$  & mas  & pc  & cm$^{-2}$ & km\,s$^{—1}$ & \\
  & \multicolumn{2}{c}{(a)}  & (b) & (c) & & & (d)  \\
    \\[-1.0ex]
    \hline \\[-1.0ex]
HZ43A & 54.106   & 84.162 & $16.6 \pm 0.05$ & $60.3 \pm 0.2$ & $14.49 \pm 0.05$ & $2.9\pm0.3$ & $-0.52\pm^{0.24}_{0.41}$	\\
PG1544+488 & 77.539   & 50.129 & $1.99 \pm 0.04$ & $497 \pm 9$  & $16.42\pm{0.07}$ &$11.1\pm{0.6}$  & $-0.45\pm{0.06} $ \\
PG1610+519 & 80.506 & 45.314 & $0.94 \pm 0.04$ & $1040 \pm^{40}_{30}$ & $16.76\pm{0.08}$ &$19.5\pm{0.8}$  & $-0.70\pm{0.09}$  \\ 
HD113001 & 110.965 & 81.163 &  $7.3 \pm 0.6$  & $140 \pm^{20}_{10}$ &$16.42\pm0.10$ &$9.7^{0.8}_{0.7}$ & $0.23\pm^{0.40}_{0.36}$ \\
Ton102 & 127.051 & 65.775 & $1.07 \pm 0.03$ & $920 \pm 30$ & $16.30\pm^{0.38}_{0.24}$ &$5.8^{2.1}_{1.5}$ & $-0.37\pm^{0.44}_{0.37}$ \\
PG1051+501 & 159.612  & 58.120 & $0.73\pm 0.04$ &  $1350 \pm^{80}_{70}$ & $16.72\pm^{0.40}_{0.26}$ &$12.6\pm^{2.6}_{2.0}$  & $0.24\pm^{0.39}_{0.32}$ \\ 
PG0952+519 & 164.068  & 49.004 & $1.70 \pm 0.05$  & $580 \pm 20$ & $16.12\pm{0.05}$ &$13.3\pm^{0.7}_{0.6}$  & $-0.54\pm{0.06}$ \\
Feige~34 & 173.315 & 58.962 & $4.36 \pm  0.10 $ & $227 \pm 5$  & $15.88\pm^{0.33}_{0.24}$ & $6.4\pm^{4.1}_{2.3}$ & $-0.40\pm^{0.43}_{0.51}$ \\
PG1032+406 & 178.877 & 59.010 & $4.67 \pm 0.05 $ & $212 \pm 2.5$ & $15.93\pm^{0.30}_{0.26}$ & $6.7\pm^{2.2}_{1.6}$  & $-0.26\pm 0.85$ \\
HD93521 & 183.140  & 62.152  & $0.78 \pm 0.08$  & $1250\pm^{150}_{100}$ & $16.46 \pm 0.08$ & - & $-0.48 \pm 0.11$  \\
     \\[-1.0ex]
    \hline
\end{tabular} 
\\[1.0ex]}
a: Galactic coordinates. \\
b: Gaia eDR3 parallax. \\
c: Median value of the distance posterior with $1\,\sigma$ uncertainty from the catalog of \citet{Bailer-Jones21}.\\
d: Abundance ratio normalized to the solar value. 
\\
\end{table*}

\begin{figure*}[!h]
\centering
\includegraphics[scale=0.47]{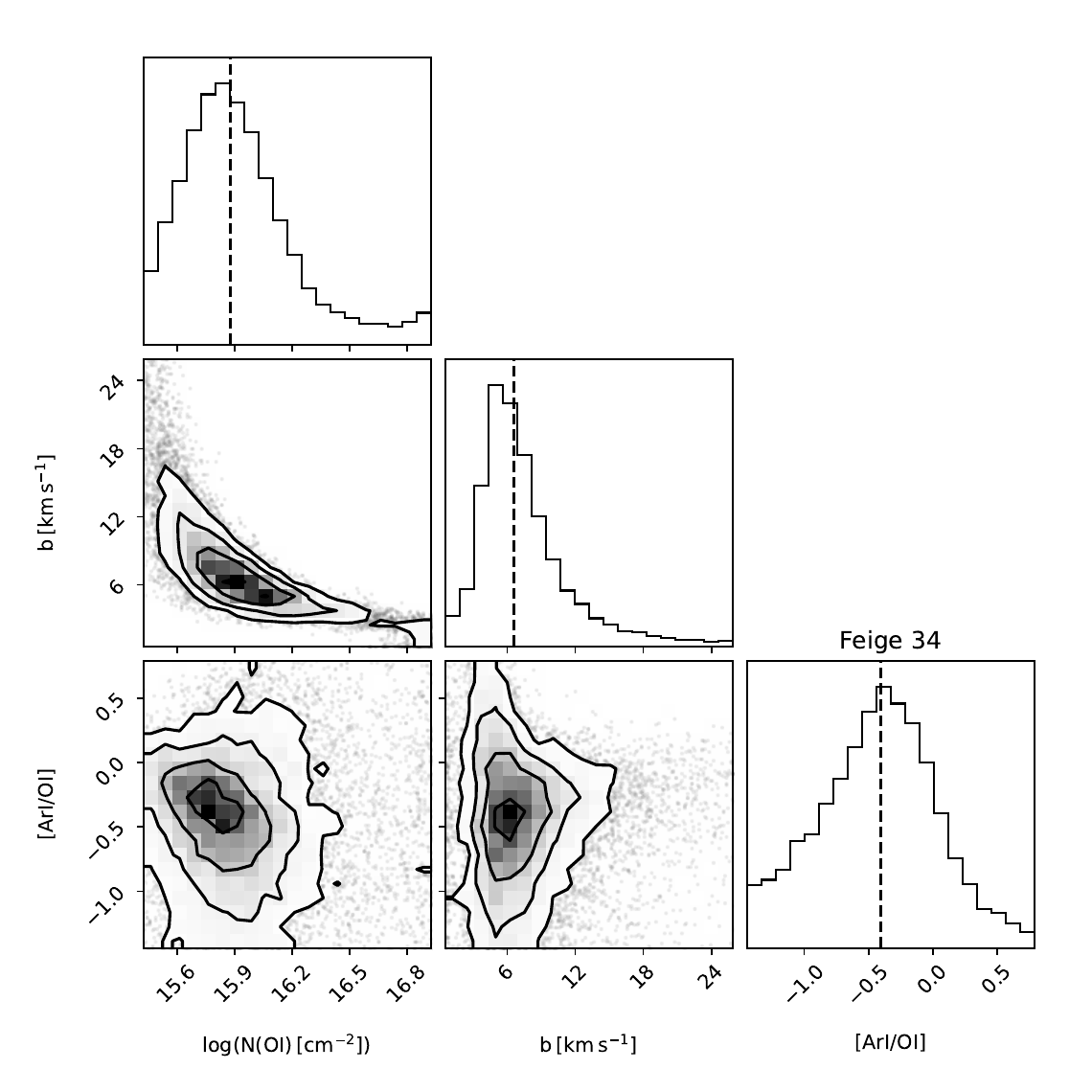}
\includegraphics[scale=0.47]{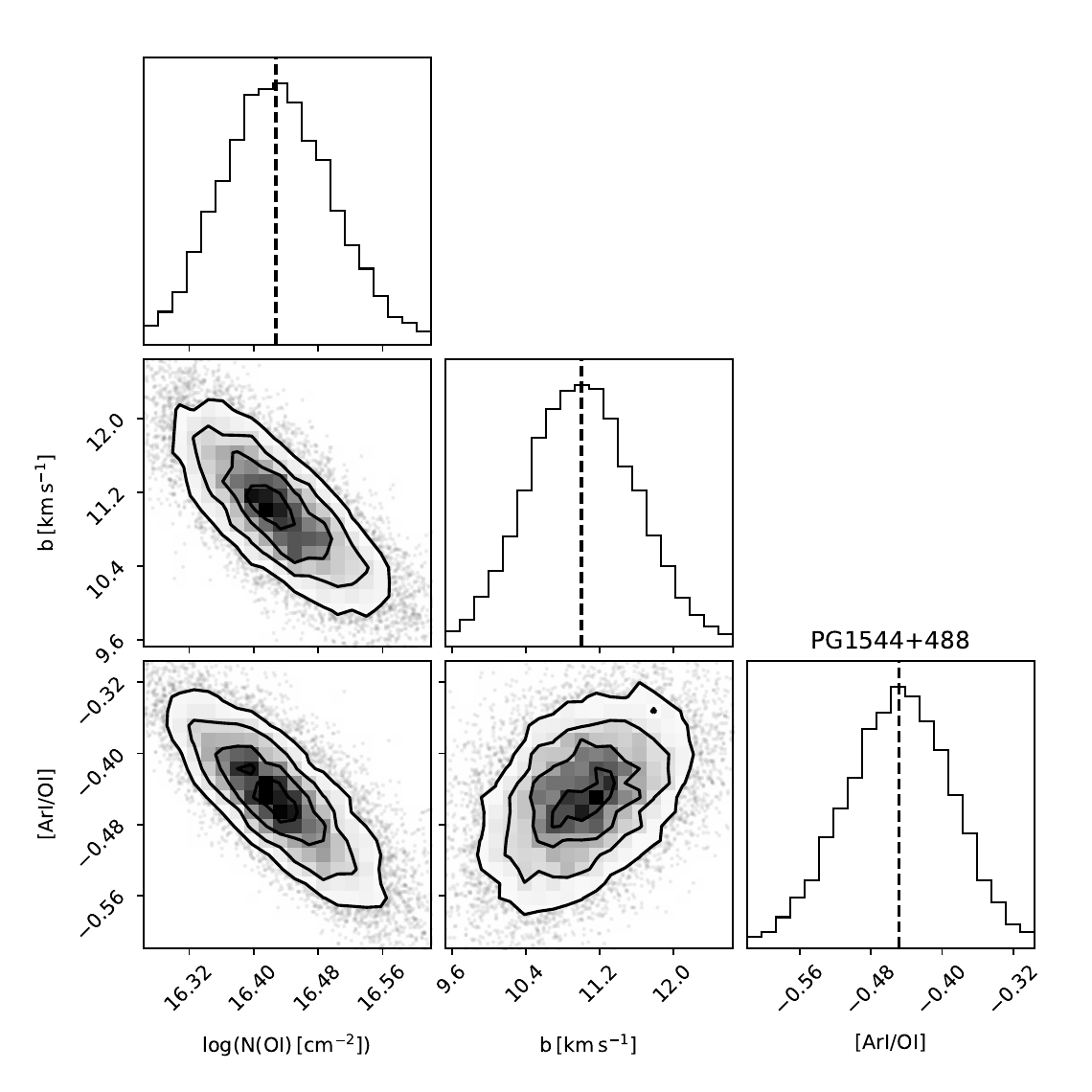}
\caption{ Corner plots of MCMC data fit  for two stars in our sample Feige~34 (left) and PG1544+488 (right), chosen as examples of low and high column densities. The histograms show the probability distribution of the three model parameters, and the contours display their interdependence. The dashed lines plotted over the histograms represent the median values of the model parameters.  }
\label{fig:corner_plots}
\end{figure*}

\begin{figure*}[!h]
\centering
\includegraphics[scale=0.47]{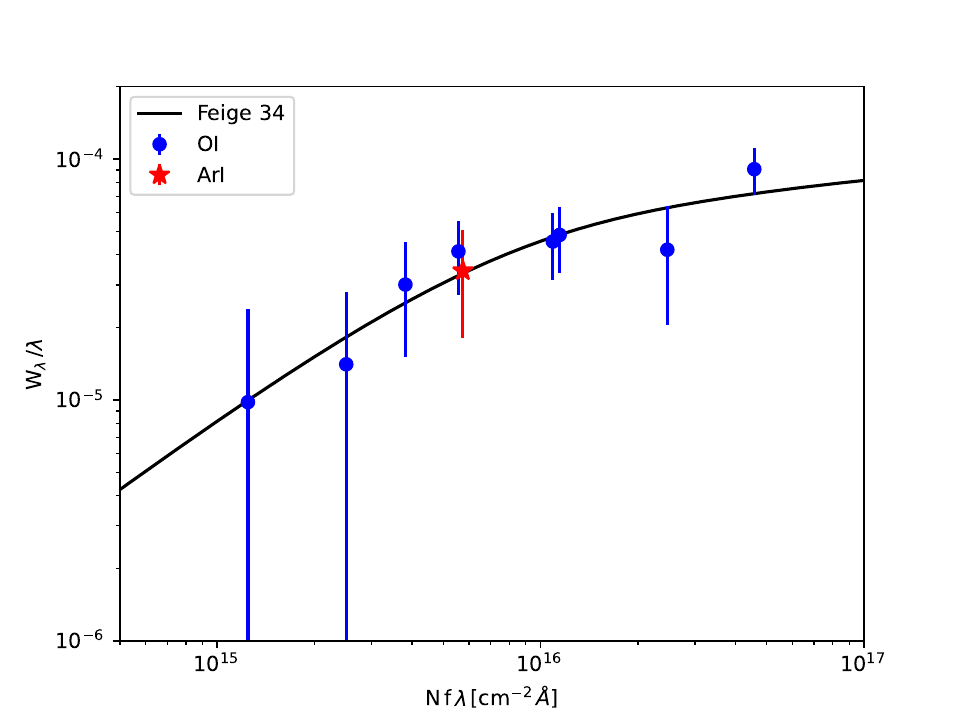}
\includegraphics[scale=0.47]{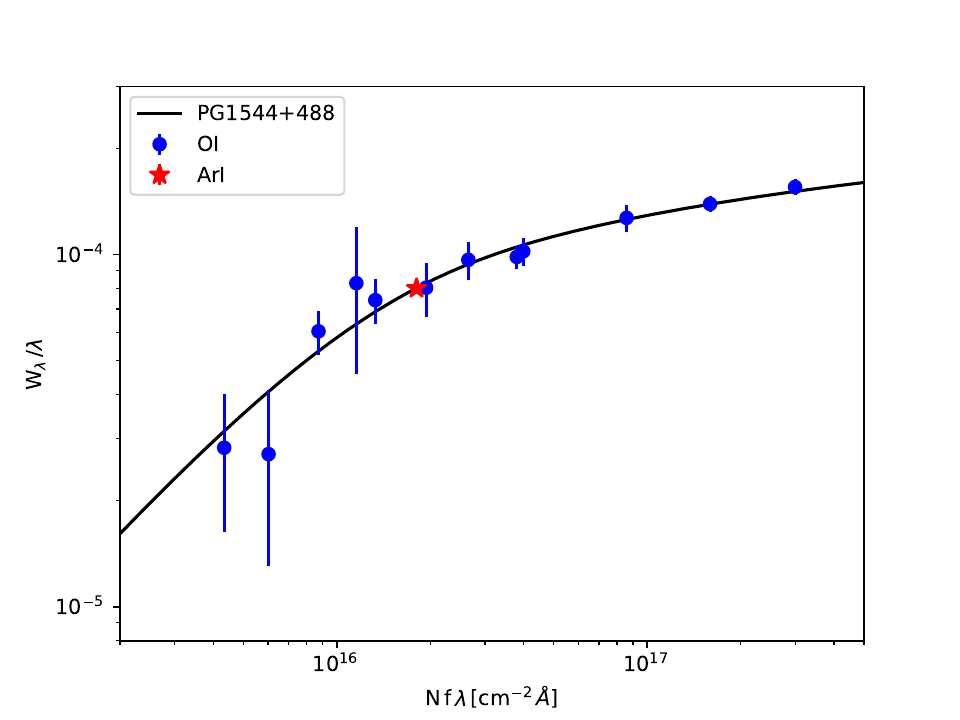}
\caption{ Curves of growth for the two stars Feige~34 (left) and PG1544+488 (right), chosen as examples of low and high column densities. For each graph, the y-axis represents the ratio between the equivalent width, W$_\lambda$, and the wavelength, $\lambda$, of the line, and the x-axis the product of the column density N expressed in cm\textsuperscript{-2}, the strength of the transition (unit-less f-value), and $\lambda $ in \AA. The blue circles represent the equivalent widths of the \OI lines used to determine the column density of \OI\ and the Doppler parameter, $b$, and the red star the equivalent width of the \ArI\ line used to determine the [\ArI/\OI] abundance ratio.}
\label{fig:OI_data_fit}
\end{figure*}

\section{Stellar targets and data}
\label{sec:stars}

We introduce our sample of stellar targets with spectroscopic UV observations in Sect.~\ref{subsec:UV_obs}. Our analysis of the UV spectra is presented in Sect.~\ref{subsec:data_analysis}. In Sect.~\ref{subsec:WNM}, the results of our data fitting are used to derive column densities of electrons in the WNM. 

\subsection{Spectroscopic UV observations}
\label{subsec:UV_obs}

To build this project, we searched scientific publications for UV spectroscopic observations, at high Galactic latitudes over the sky area covered by the LOFAR Faraday tomography mosaic (hereafter the LoTSS field) analyzed by \citet{Erceg22}. 
We obtained a sample of nine stellar sources. This sample includes eight stars from the study of \citet{Jenkins13} plus the halo star HD~93521 \citep{Spitzer93}. To this set, we added the star HZ43A \citep{Kruk02}, which lies close to the edge of the LoTSS field. This star, located $60\,$pc from the Sun, allows us to probe the contribution of the very nearby gas in this region of the sky.  

In Fig.~\ref{fig:LOFAR_map}, the stellar positions are overlaid on an image of the first moment of the Faraday data cube from \citet{Erceg22}, representing the polarized intensity-weighted mean of Faraday spectra in units of ${\rm rad~m^{-2}}$. 
All stars have been 
observed with the Far Ultraviolet Spectroscopic Explorer (FUSE) with a spectral resolution $R\equiv \nu/\Delta \nu \simeq 2\times 10^4$. 

Stellar coordinates and data relevant to this study are listed in Table~\ref{stellar_data}.  
The parallaxes and corresponding distances are from the Gaia data release eDR3 \citep[][]{Lindegren21}.  
The column densities of \ArI\ and \OI\ are inferred from the FUSE UV spectra. 
We follow \cite{Jenkins13} using the [\ArI/\OI] column density ratio, normalized to Solar abundances, to estimate the ionization fraction of the WNM. The ionization potentials of \ArI\ and \HI\ are close, but the photo-ionization cross section of \ArI\ is much larger. The ionization fraction of \OI\ is strongly locked to that of \HI\ through a strong charge-exchange reaction. These two factors make [\ArI/\OI] a sensitive tracer of \HI\ ionization in the WNM \citep{Sofia98}.
Since neutral forms of argon and oxygen are virtually absent in fully ionized regions (either the prominent \HII\ regions around hot stars or the much lower density but more pervasive WIM), 
our probes sample only regions that have appreciable concentrations of \HI. This 
sets our measurements apart from conventional determinations of average electron densities (e.g. pulsar dispersion measures, H$_\alpha$ line intensity, \CII\ fine-structure excitation), which are strongly influenced by contributions from fully ionized gas.

\subsection{Data analysis}
\label{subsec:data_analysis}

The column densities of \OI\ and \ArI\ (number of absorbing atoms in a column of unit cross-section along the LoS) are determined by a curve of growth analysis \citep{Stromgren48,Draine11}. The curve of growth relates the gas column densities of the absorbers, $N$, to the strength of absorption lines measured by their equivalent width $W_\lambda$ (normalized area of the line expressed in wavelength units). The function $W_\lambda (N)$ is computed assuming that the absorbers have a Gaussian velocity distribution characterized by the Doppler width $b = \sqrt{2} \times \sigma_v$, where $\sigma_v$ is the velocity dispersion. By fitting the equivalent widths of multiple lines of different strength one can infer the two parameters $N$ and $b$ from the data.  

We used the equivalent widths of multiple \OI\ lines and a single \ArI\ line, measured by \citet{Jenkins13}, and by \citet{Kruk02} for HZ43A. The oscillator strengths of the lines are listed in these two papers. 
The fit of the curve of growth is performed with three free parameters: the column density $N(\OI)$ of \OI\ atoms, the abundance ratio [\ArI/\OI] normalized to the Solar value, and the Doppler parameter, $b$, of the line profiles, assumed to be the same for \ArI\ and \OI. 
The spontaneous decay rate that determines the Lorentzian broadening is set to $A = 3\times 10^7\,$s$^{-1}$ ignoring variations among \OI\ states \citep{Morton03}. This simplification has no significant impact on the results of our data analysis.

We performed a Monte-Carlo Markov Chain (MCMC) analysis to sample the joint posterior distribution
of the three model parameters. We utilize the EMCEE python package as described by \citet{EMCEE13} in accordance with their recommendations.
We use a Gaussian likelihood and broad priors on the model parameters. We restricted $b$ to values greater than 2\,km\,s$^{-1}$ 
 because we consider that in our LoS the \OI\ lines are dominated by warm gas (the partially ionized WNM), which implies a temperature of T $\geq$ 6000 K and, thus, a line broadening $b > 2 $ km/s. To confirm this hypothesis   
we looked at the \HI\ 21\,cm spectra in direction of each of our stars using the HI4PI data from \citet{HI4Pi16} and their Gaussian decomposition by \citet{Kalberla18}. These data indicate that some of the gas along our LoS is cold but its contribution to the OI absorption lines is minor due to line saturation. The cold neutral medium (CNM) could contribute to very weak transitions but the \OI\ 
lines that have the most influence in our analysis have equivalent widths that are 
not much different from that of the \ArI\ line.  In this case, comparisons of \ArI\ 
to \OI\ remain valid, and the very narrow components from the CNM have no influence.  
Moreover, fits to the data for small values of $b$ suppose that the lines are highly saturated and give \OI\ column densities that are incompatible with \HI\ column densities inferred from \HI\ $21\,$cm observations. 
Figure~\ref{fig:corner_plots} presents the posterior
distributions of the model parameters for two stars Feige~34 and PG1544+488 chosen as examples of low and high column densities of \OI. The corresponding data fits with the curves of growth are shown in Fig.~\ref{fig:OI_data_fit}. 

The model parameters with their error bars are listed in Table~\ref{stellar_data}.
The best estimate is the median value of the posterior distribution, and the negative and positive error bars are computed from the 16 and 84 percentiles. We added to Table~\ref{stellar_data} data for the star HD~93521. Interstellar matter foreground to HD~93521 has been analyzed in detail by \citet{Spitzer93}. 
Here, we report the \OI\ column density and the [\ArI/\OI] abundance ratio that we measured in the FUSE spectra for the low-velocity component. We checked that our estimates of [\ArI/\OI] are consistent with the values reported by \citet{Jenkins13} but for PG1610+519\footnote{ For this star, the linear fit used by \citet{Jenkins13} departs from the curve of growth because only a few \OI\ lines were measured.}. 
For this parameter of the model, the added value of our data analysis is the posterior distribution that quantifies the interdependence with $N(\OI)$ and $b$.

\begin{table}

{\small
    \caption{Column densities and ionization fractions}
\label{tab:column_WNM}
   \begin{tabular}{llll}
    \hline
    \hline
    \\[-1.0ex]
 Star    & log$(N({\rm HI}))$ & log$(N^{\rm WNM}_e)$  & log$(x_H)$ \\
        & cm$^{-2}$ &cm$^{-2}$ & \\
   & (a) & (b) & (c)  \\
    \\[-1.0ex]
    \hline \\[-1.0ex]
HZ43A & $17.73\pm{0.05}$  & $16.75 \pm_{0.42}^{0.47} $ & $0.10 \pm_{0.06}^{0.19} $  \\
PG1544+488 & $19.66\pm {0.07}$ & $18.62\pm{0.17}$ &$0.090\pm_{0.020}^{0.024} $\\
PG1610+519 & $20.0\pm {0.08}$ & $19.36\pm {0.21}$ & $0.23\pm_{0.06}^{0.09}$ \\
HD113001 & $19.65\pm{0.10}$ & $<18.2 $ &$< 0.035$ \\
Ton102 & $19.52\pm_{0.23}^{0.34}$   &  $18.35\pm_{1.09}^{0.71}$ & $0.072\pm_{0.065}^{0.165}$ \\
PG1051+501 & $19.93\pm_{0.25}^{0.38}$ & $< 18.35$ & $< 0.021$ \\
PG0952+519 & $19.36\pm {0.05}$  & $18.46\pm{0.14}$ & $0.126\pm_{0.024}^{0.03} $\\
Feige~34 & $19.11\pm_{0.23}^{0.31}$  & $17.95\pm_{1.56}^{0.76}$ & $0.069\pm_{0.066}^{0.215}$ \\
PG1032+406 & $19.17	\pm_{0.25}^{0.29}$	& $17.98 \pm_{0.94}^{0.82}$ & $0.066\pm_{0.076}^{0.324}$ \\
HD93521 & $19.70 \pm 0.08$ & $18.71 \pm_{0.21}^{0.19}$ & $0.102 \pm_{0.036}^{0.051}$  \\
     \\[-1.0ex]
    \hline
\end{tabular} 
\\[1.0ex]}
(a): Log$_{10}$ of the column density of HI derived from that of OI. \\
(b): Log$_{10}$ of the column density of electrons.  \\
(c): Hydrogen ionization fraction. 
\end{table}

\subsection{Column densities of electrons}
\label{subsec:WNM}

The column densities of electrons for the WNM, $ N_e^\mathrm{WNM}$ foreground to the stars are estimated from the results of our fits of FUSE spectra. We follow the approach introduced by \citet{Jenkins13}, to derive the ionization fraction of HI, $x_{\rm H} \equiv n({\rm H}^+)/n({ \HI})$, from the [\ArI/\HI] abundance ratio. We use formula (7) in this paper:
\begin{equation}
    [{ \ArI/\HI }] = {\rm log}  \left[\frac{1+ x_{\rm H}}{1+P^\prime_{\rm Ar}\, x_{\rm H}}\right],
\label{eq:ArI_HI}
\end{equation}
where $P^\prime_{\rm Ar}$ is a unit-less parameter, which accounts for various forms of photoionization and recombination with free electrons, as well as additional processes \citep{Jenkins13}. 
[\ArI/\HI] is deduced from [\ArI/\OI] for the Solar abundance of oxygen $\rm log[O/H] + 12 = 8.76$. We assume that there is no depletion of oxygen on dust grains, which should be safe for the low-density conditions we are sampling here \citep{Jenkins09}.
We use
$P^\prime_{\rm Ar} = 22.8$, the value from Table~4 in  \citet{Jenkins13} for a model including ionization by soft X-ray emission from cooling supernova remnants. The statistical distribution of $x_{\rm H}$ is derived
from the posterior distribution of [\ArI/\OI] using Eq.~\ref{eq:ArI_HI}, and that of the \HI\ column density $N({\HI})$  from $N({\OI})$ . 

We combine $x_{\rm H}$ and $N({\HI})$ values to compute the electron column density:
\begin{equation}
    N_e^{\rm WNM} = 1.2 \, x_{\rm H} \, N({\HI}),
\label{eq:Nelec}
\end{equation}
where the factor 1.2 accounts for electrons from ionized helium, based on the ionization model of the WNM in \citet{Jenkins13}. 

\begin{figure}[!h]
\centering
\includegraphics[scale=0.6]{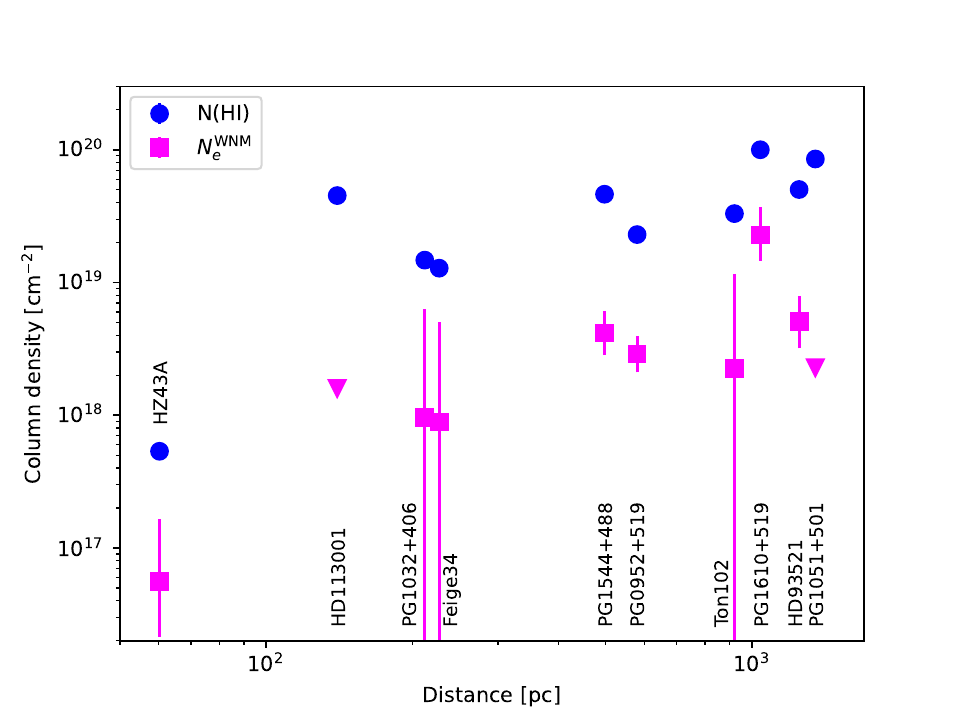}
\caption{Column densities of HI and WNM electrons. The HI column densities are plotted with blue circles and the electron column densities in the WNM with magenta squares. The two triangles pointing down represent upper limits. 
}
\label{fig:column_densities}
\end{figure}

The column densities of neutral hydrogen and electrons are listed in Table~\ref{tab:column_WNM} with their error bars and plotted versus distance in Fig.~\ref{fig:column_densities}.  $N_e^{\rm WNM}$  and  $N(\HI)$ seem to be correlated : <$N(\HI)$ / $N_e^{\rm WNM}$ > = 10.9 with a small dispersion of 3.6.  We observe a tendency for $N$(\HI)  to increase with distance but our star sample is too small to draw some firm conclusion about the distance distribution of WNM electrons.

\begin{figure*}[!h]
\centering
\includegraphics[scale=0.55]{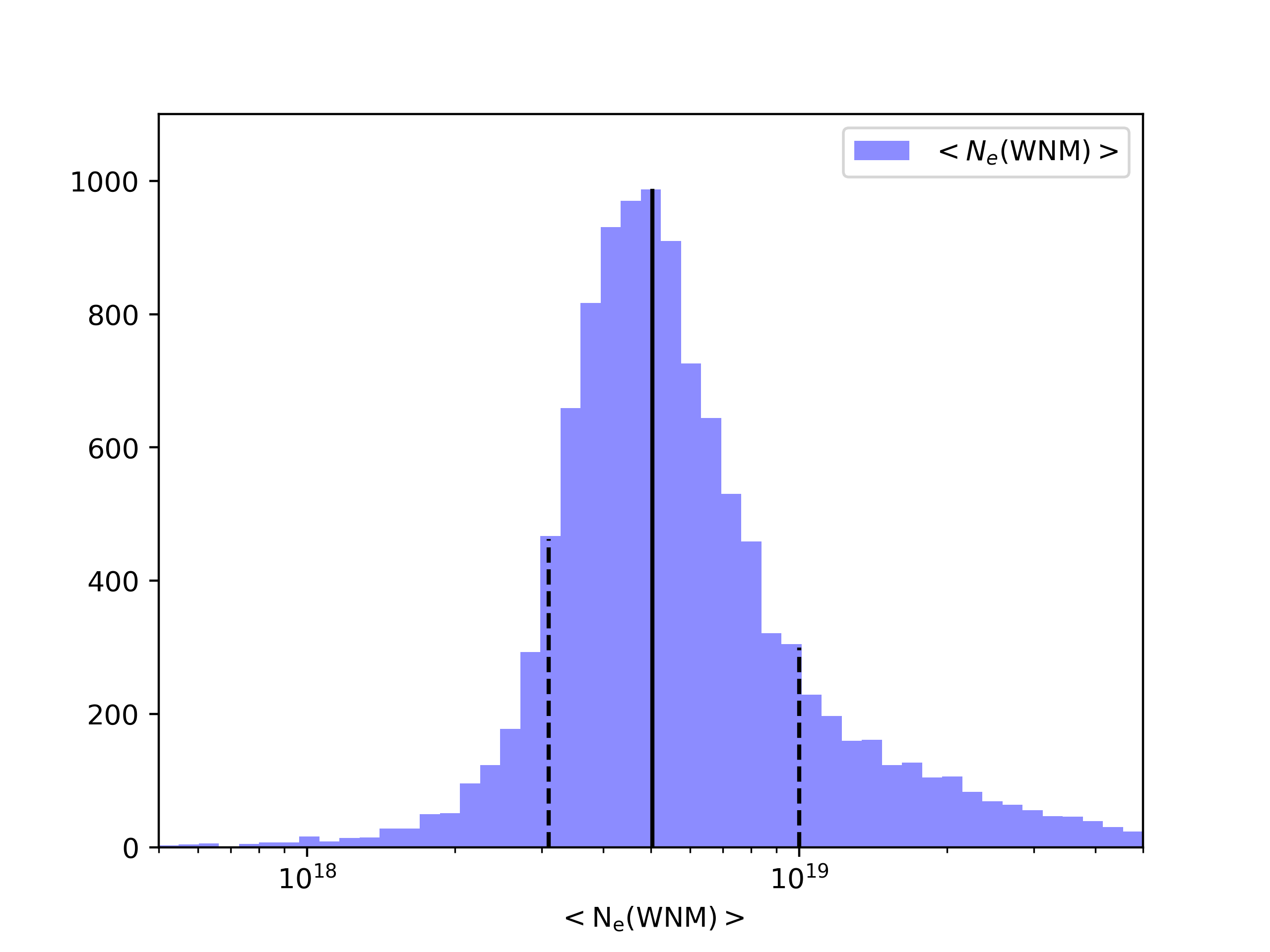}
\includegraphics[scale=0.55]{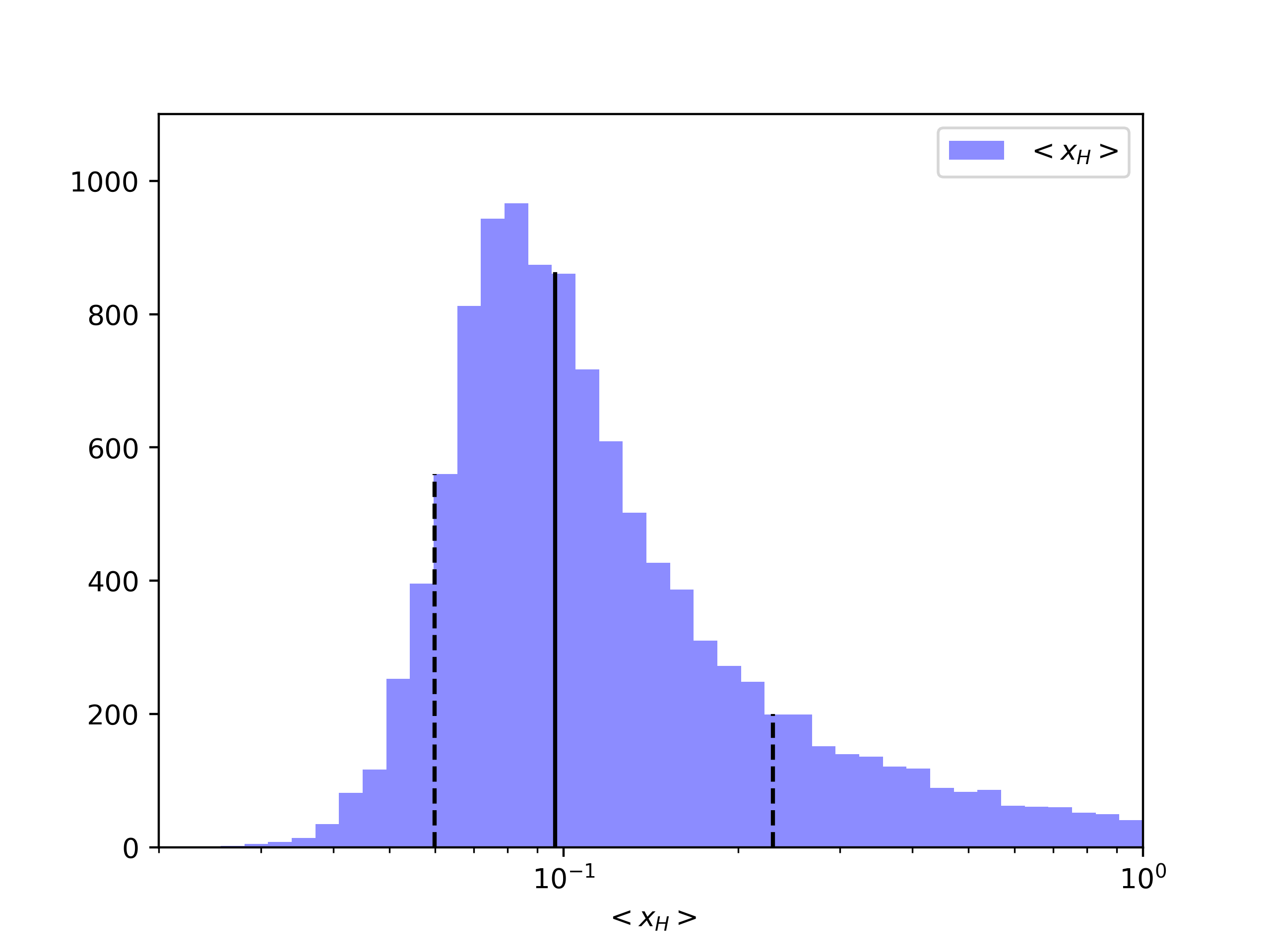}
\caption{Probability distribution of the mean value of the electron column density (left) and ionization fraction (right). The histograms are built averaging the samples of the posterior distributions over the 9 stars in our sample more distant than 100\,pc (i.e. excluding HZ43A). The solid black lines indicate the median value and the dashed lines the 16 and 84\% percentiles.   
%horizontal magenta strip shows the mean column density $N_e^{\rm Shell}$ with its error-bar, and the
%The vertical grey strip the range of distances to the edge of the LB for our sample of sightlines. 
}
\label{fig:pdf_x_H}
\end{figure*}

The probability distribution functions (PDFs) of the mean column density of electrons and the ionization fraction averaged over our set of stars are presented in Fig.~\ref{fig:pdf_x_H}. The PDFs are computed by averaging the samples of the posterior distributions, inferred from our data analysis of UV spectra, over the nine stars in Table~\ref{tab:column_WNM} more distant than HZ43A. The nearest star in this set is located $140\,$pc from the Sun. The median values for the electron column density and the ionization fraction are $\langle  N_e^\mathrm{WNM} \rangle = 5\pm^5_2\times 10^{18}\, \mathrm{cm}^{-2}$ and $\langle x_{\rm H}\rangle = 0.097\pm^{0.133}_{0.037}$, where the lower and upper error bars represent the 16 and 84 percentiles of the distributions. Our median value is consistent with that inferred by \citet{Jenkins13} from the complete sample of his study. For $P^\prime_{\rm Ar} = 13.4$, the model value without the contribution of cooling supernovae \citep{Jenkins13}, we find $\langle x_{\rm H}\rangle = 0.17$. The difference quantifies the dependence of our $x_{\rm H}$ estimates on the choice of ionization model. 

The various sources and ionization processes that contribute to the ionization fraction of the WNM are detailed by \citet{Jenkins13}. These include cosmic rays as well as supernova-driven shocks \citep{Sutherland:2017} and soft X-ray emission from cooling supernova remnants \citep[SNRs,][]{Slavin00}. The degree of ionization of the WNM could also be locally out of equilibrium as an aftereffect of the supernova explosions that have created the LB \citep{Fuchs06,Zucker22}. Within this hypothesis, the WNM ionization could be enhanced around the LB. Analysis of UV data for a larger sample of stars is required to test this hypothesis.

\section{Model of the local magnetic field from \Planck\ observations} 
\label{sec:Bfield_model}

To estimate the RM from the partially ionized WNM foreground to the stars,
we need a model of the magnetic field in the local ISM. The tight correlation between the orientation of Faraday structures in LOFAR data with that of dust polarization \citep{Zaroubi:2015,Jelic:2018} leads us to use the phenomenological model introduced by \citet{PIPXLIV} and \citet{Vansyngel17} to statistically model dust polarization data at high Galactic latitudes as measured by \Planck.  

The magnetic field $\vec{B}$ is
expressed as the sum of its mean (ordered) $\vec{B_0}$ and random (turbulent) $\vec{B_{\rm t}}$ components:
\begin{equation}
\vec{B} = \vec{B_0} + \vec{B_{\rm t}} = |\vec{B_0}| \, ( \vec{\hat{B}_0} + f_{\rm M} \, \vec{\hat{B}_{\rm t}}),
\label{eq:Bmodel}
\end{equation}
where $\vec{\hat{B}_0}$ and $\vec{\hat{B}_{\rm t}}$ are the unit vectors associated with $\vec{B_0}$ and $\vec{B_{\rm t}}$,
and $f_M$ a model parameter that sets the relative strength of the random component of the field. The direction of $\vec{B_0}$ is assumed to be uniform over the sky area used to fit the model parameters. The model also makes the simplifying assumption that the direction of $\vec{B_0}$ does not vary along the LoS. At high Galactic latitudes ($|b| \ge 60^\circ$), comparison with stellar polarization data indicates that dust polarization measured by \Planck\ originates mainly from the surroundings of the LB, within about $300\,$pc from the Sun \citep{Skalidis19}. Even if the LOFAR field in Fig.~\ref{fig:LOFAR_map} extends to lower latitudes, it is not necessary to account for the structure of the magnetic field 
across the Galactic disk and into the halo to model \Planck\ dust polarization data over this sky area. 

The computational steps taken to determine the magnetic field model are detailed in Appendix~\ref{App:Bfield}.
The direction of $\vec{\hat{B}_0}$ is defined by the
Galactic coordinates $l_0$ and $b_0$, which are derived from a fit of the \Planck\ Stokes $Q$ and $U$ maps at $353\,$ GHz over the sky regions toward the northern Galactic cap comprising the LoTSS field. The random component $\vec{B_{\rm t}}$ is statistically described using the model of \citet{PIPXLIV} with parameters derived from a fit of the \Planck\ power spectra of dust polarization by \citet{Vansyngel17}. The orientation of $\vec{B_{\rm t}}$ varies in discrete steps along the LoS over a small number of layers, which schematically represent the correlation length of the turbulent component of the magnetic field. The ratio between $|\vec{B_{\rm t}}|$ and $|\vec{B_0}|$,  $f_M = 0.9$, is determined fitting \Planck\ power spectra of dust polarization \citep{Vansyngel17}.

Because the modeling of dust polarization only determines the  orientation of the mean magnetic field, we need to use  Faraday observations to determine the field strength and also to discriminate between the two opposite directions associated with the orientation\footnote{We select the $l_0$ values in Table ~\ref{tab:B0_fit} to be between 0 and $90^\circ$}. Comparison of the rotation and dispersion measures of pulsars in the northern sky indicates that the mean field strength in diffuse ionized gas in the Galactic disk near the Sun is about 3-$4\,\mu$G \citep[see Table~3 in][]{Sobey19}. The magnetic field strength in the warm interstellar medium can also be estimated from synchrotron emission. As discussed in \citet{Beck03}, the two estimates  differ if fluctuations in magnetic field and electron density are not statistically independent. For the specific objective of the paper, the strength of the mean field estimated from pulsar RMs is a directly relevant reference. We chose this strength $|\vec{B_0}|$ to be ${\bf 3.5}\,\mu$G. The total field strength that includes the turbulent component is $ 5\,\mu$G.

\begin{figure}[!h]
\centering
\includegraphics[scale=0.26]{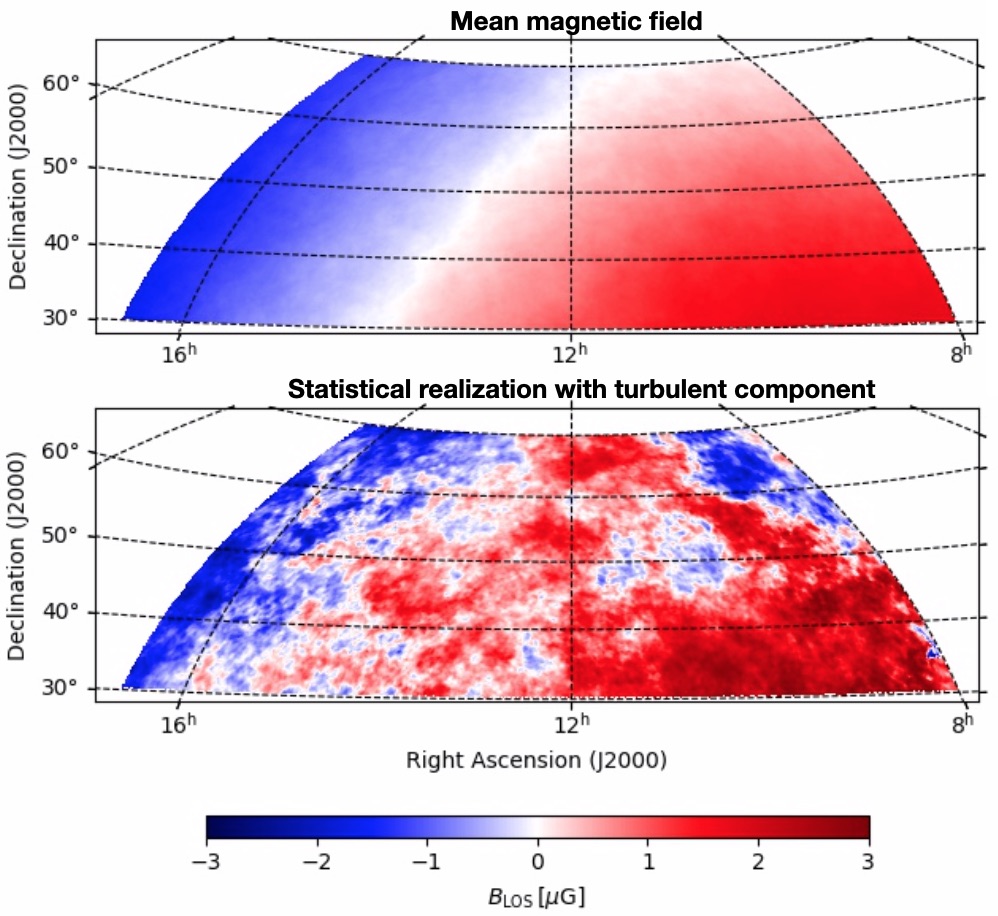}
\caption{ Maps of the magnetic field component along the LoS, $B_\mathrm{LoS}$, over the LOFAR field. The top image shows $B_\mathrm{LoS}$ for the mean field $\vec{B_0}$ and the bottom image that computed for $\vec{B}$ including one realization of the turbulent component $\vec{B_{\rm t}}$.  $B_\mathrm{LoS}$ is positive when it points toward the Sun (See Eq.~\ref{eq:BLOS}).  }
\label{fig:Bfield}
\end{figure}

To estimate the RM from electrons in the WNM, we compute the mean magnetic field component along the LoS, $B_\mathrm{LoS} $:
\begin{equation}
B_\mathrm{LoS} = -\, (\vec{B_0} + <\vec{B_{\rm t}}>)\cdot \hat{r},
\label{eq:BLOS}
\end{equation}
where $\vec{B_{\rm t}}$ is averaged over the model layers and $\hat{r}$  is the unit vector defining the LoS.  The minus sign makes $B_\mathrm{LoS}$ positive when it points toward the Sun.
Figure~\ref{fig:Bfield} presents our calculation results in the LOFAR field. The top image shows the mean value of $B_\mathrm{LoS}$. This map represents $B_\mathrm{LoS}$ computed for the mean magnetic field $\vec{B_0}$, ignoring the random component. 
The bottom image shows $B_\mathrm{LoS}$ for one statistical realization of $\vec{B_\mathrm{t}}$. The mean value of $B_\mathrm{LoS}$ and the dispersion associated with $\vec{B_{\rm t}}$ are listed in Table~\ref{tab:Faraday_depth}. 

\begin{table}[ht!]
{\small
\caption{$B_\mathrm{LoS}$  and WNM RMs}       
\label{tab:Faraday_depth}      
\begin{tabular}{ lccc}    
\hline\hline  
\\[-1.0ex]
   Star & $B_\mathrm{LoS}$ & RM$_\mathrm{WNM}$ & $ \mathrm{M1_{LOFAR}}$    \\  
       & $\mathrm{\mu G}$  & rad\,m$^{-2}$  & rad\,m$^{-2}$    \\
        & (a)  & (b) & (c)  \\

    \\[-1.0ex]
   \hline 
   \\[-1.0ex]
    HZ43A  & $0.49 \pm 1.0$ & $0.0\pm ^{0.02}_{0.01}$ & ~~~$-$  \\ 
    PG1544+488  & $-1.3 \pm 0.9$ & $-1.4\pm^{1.0}_{1.4}$ & $-1.39  $ \\
    PG1610+519 & $-1.5 \pm 0.9$ & $-8.6 \pm^{5.8}_{9.3}$ & $0.88 $ \\
    HD113001 & $0.44 \pm 1.0$ & $0.0\pm {0.2}$  & $-0.32 $ \\
    Ton 102  & $0.16 \pm 1.0$ & $0.0 \pm^{1.0}_{0.6}$ & $-0.67  $ \\ 
    PG1051+501 & $0.82 \pm 1.0$ & $0.0 \pm {0.5} $ & $-0.46 $ \\
    PG0952+519  & $1.0 \pm 0.9$ & $0.6 \pm^{0.8}_{0.7}$ & $-2.02  $ \\ 
    Feige~34  & $1.2 \pm 0.9$ & $0.2 \pm^{1.4}_{0.2} $ & $ 1.00 $ \\
    PG1032+406 & $1.4 \pm 0.9$ & $0.2 \pm^{2.0}_{0.3} $ & $ 0.0 $ \\
    HD93521  & $1.6 \pm 0.9$ & $1.7\pm ^{2.1}_{1.3}$ & ~~~$-$  \\
    \\[-1.0ex]
    \hline
\end{tabular} 
\\[1.0ex]}
(a) Mean $B_\mathrm{LoS}$ inferred from the \Planck\ magnetic field model followed by the standard deviation of the turbulent component B$_t$.\\
(b) RM to the star associated with the foreground WNM.\\
(c) First moment of the LOFAR Faraday spectra. The uncertainty of these values is $1\,\mathrm{rad\,m^{-2}}$. 
\end{table}

\section{The warm neutral medium and LOFAR Faraday observations}
\label{sec:LOFAR}

In this section, we assess whether the partially ionized WNM foreground to the stars may account for the LOFAR Faraday spectra.  The RMs associated with electrons within the WNM foreground to the stars are determined in Sect.~\ref{subsec:RMs}, and compared with the LOFAR data in Sect.~\ref{subsec:comp_LOFAR}. Electrons in the WIM are ignored in this section.

\begin{figure}[!h]
\centering
\includegraphics[scale=0.57]{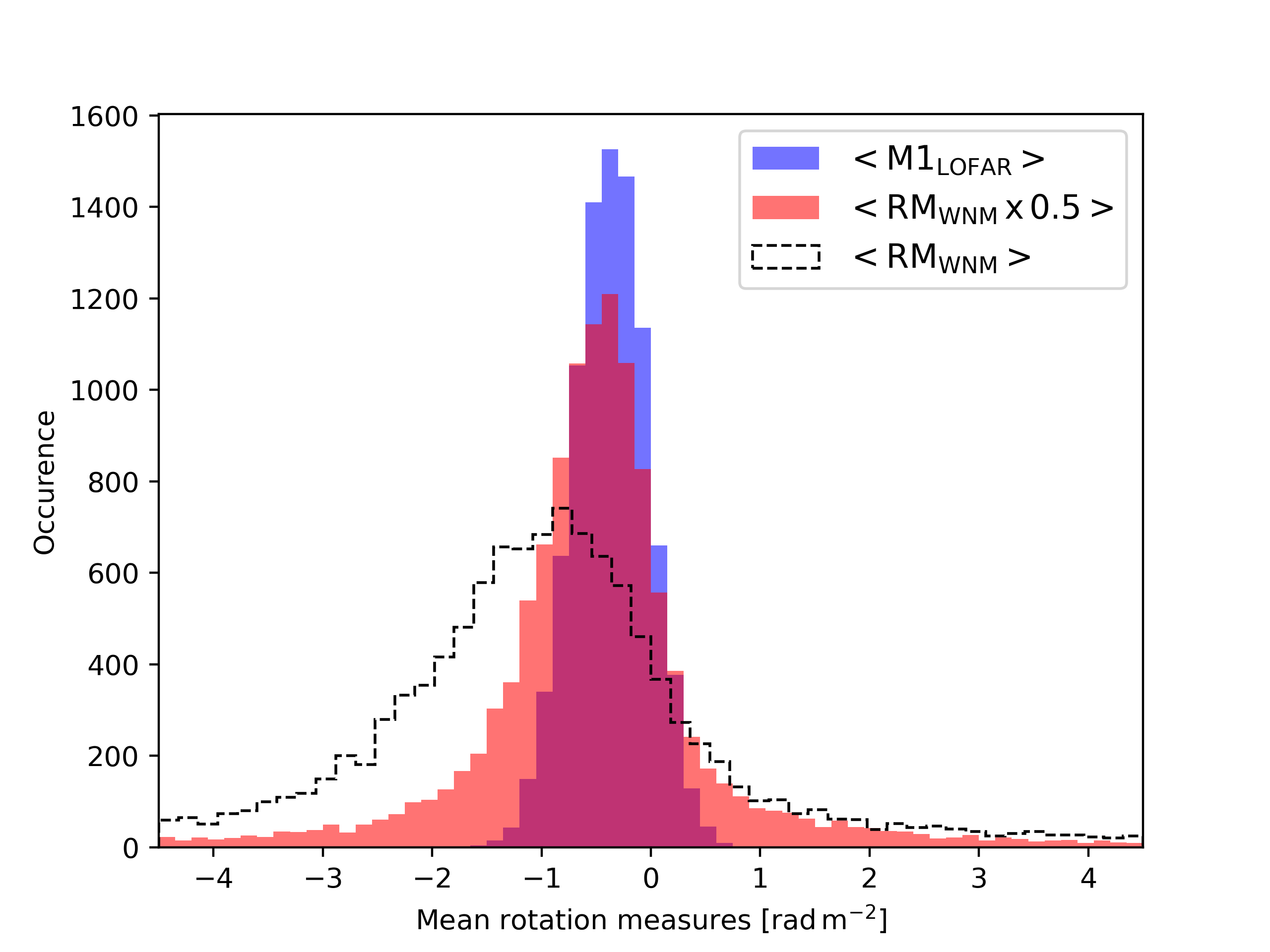}
\caption{PDFs of the mean RM and the mean LOFAR M1 moment. The plot displays the PDFs of $\langle \mathrm{RM_{WNM}} \times 0.5 \rangle$ (red) and $\langle M1_\mathrm{LOFAR} \rangle$ (blue). The PDF of $\langle \mathrm{RM_{WNM}} \rangle$ is also shown as a black line. The mean values are computed over the eight stars with both UV and LOFAR data.  }
\label{fig:Faraday_mean_histo}
\end{figure}

\subsection{Rotation measures from WNM electrons}
\label{subsec:RMs}

The RM to a source at a distance $d$ is defined as
\begin{equation}
\mathrm{RM}(d) = 0.812\,{\rm rad\,m^{-2}} 
\int^{d}_{\rm{0}} \left(\frac{n_{\rm e}}{\rm cm^{-3}}\right)  
\left(\frac{B_{\parallel}}{\mu{\rm G}}\right)  
\left(\frac{dl}{\rm pc}\right),
\label{eq:phi}
\end{equation}
where the integral along the LoS goes from the source at $l=0$ to the observer at $l=d$, $n_e$ is the electron density and $B_{\parallel}$ the magnetic field component along the LoS \footnote{ $B_{\parallel}$ is positive when it points toward the Sun.}. Using $B_\mathrm{LoS}$, the mean value of $B_{\parallel}$ introduced in Eq.~\ref{eq:BLOS}, 
we reduce the integral in Eq.~\ref{eq:phi} to 
\begin{equation}
\mathrm{RM_{WNM}}(i) = 0.26\,{\rm rad\,m^{-2}} \, 
 \left(\frac{N_{\rm e}^{\rm WNM}(i)}{\rm 10^{18}\, cm^{-2}}\right) \,  
\left(\frac{B_\mathrm{LoS}(i)}{\mu{\rm G}}\right),
\label{eq:phi_star}
\end{equation}
where $\mathrm{RM_{WNM}}(i)$ is the RM associated with the column density of electrons $N_{\rm e}^{\rm WNM}(i)$ to the star $i$.  
Within our statistical model of the magnetic field (Sect.~\ref{sec:Bfield_model}), $B_\mathrm{LoS}$ for a given LoS has a normal distribution. The mean values and standard deviations of $B_\mathrm{LoS}$  are listed for the LoS to the stars in Table~\ref{tab:Faraday_depth}. The PDF of $B_\mathrm{LoS}$ is combined with the posterior distribution of $N_{\rm e}^{\rm WNM}$, inferred from our analysis of the FUSE spectra, to obtain the statistical distribution of $\mathrm{RM_{WNM}}$. The median values of $\mathrm{RM_{WNM}}$ with $1\sigma$ uncertainties  are listed in Table~\ref{tab:Faraday_depth}. These uncertainties are large for all stars because they include the statistical dispersion of the turbulent component of $B_\mathrm{LoS}$.

\begin{figure}[!h]
\centering
\includegraphics[scale=0.6]{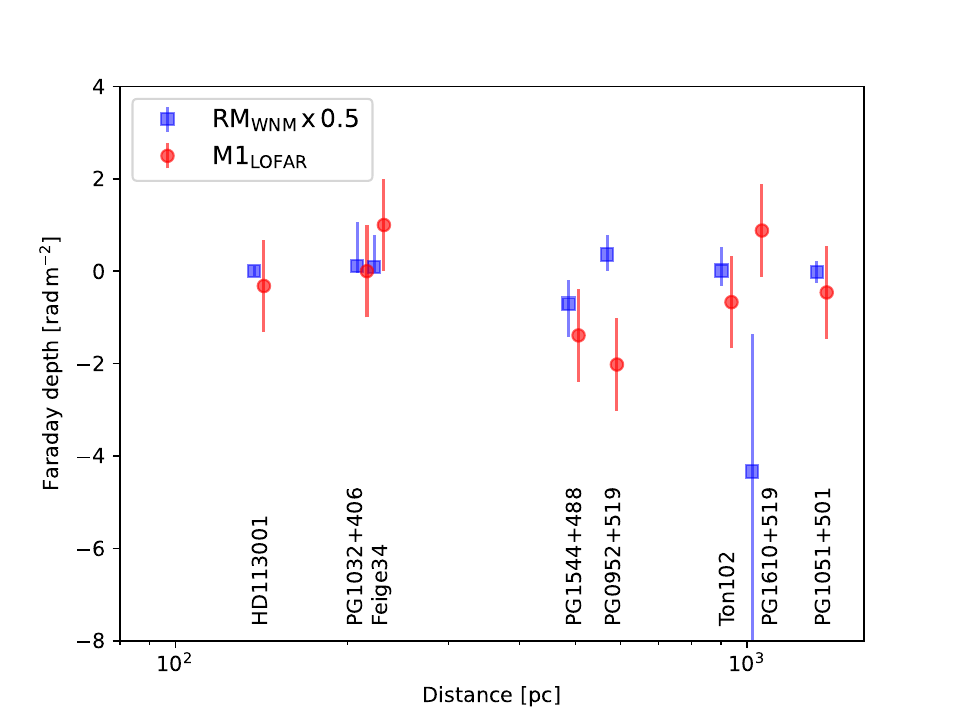}
\caption{  
Comparison of the WNM RMs with the first moments of  the LOFAR Faraday spectra for each star individually. The values of $\mathrm{RM_{WNM}} \times 0.5$ are plotted with blue squares, and the intensity weighted Faraday depths from LOFAR, $M1_\mathrm{LOFAR}$, with red circles. 
}
\label{fig:RMstar_vs_distance}
\end{figure}

\subsection{Comparison with LOFAR data}
\label{subsec:comp_LOFAR}

We now proceed to the comparison between $\mathrm{RM_{WNM}}$ and the LOFAR data.

The third column of Table~\ref{tab:Faraday_depth} lists the first moment of the Faraday spectra for each LoS to the stars. The first moment $\mathrm{M1_{LOFAR}}$ is the mean in Faraday depth of the spectra, weighted by the polarized intensity.
The $\mathrm{M1_{LOFAR}}$ values were computed by \citet{Erceg22} for a polarized intensity threshold of $460~{\rm \mu Jy}$ per angular and Faraday resolution elements. No values could be measured for HD~93521 because the polarized emission at this position is too weak to be detected. HZ43A is just outside the field of the LoTSS field in Fig.~\ref{fig:LOFAR_map}. 

To reduce the uncertainties in the observational data and the statistical dispersion of the turbulent component of the magnetic field, we compared the mean values  of $\mathrm{RM}_{WNM}$ and $\mathrm{M1_{LOFAR}}$, computed for the eight stars with the LOFAR values in Table~\ref{tab:Faraday_depth}. 
The PDFs of $\langle \mathrm{RM}_{WNM} \rangle$ are computed by combining the statistical distributions of $N_e^{\rm WNM}$ and $B_\mathrm{LoS}$. The PDF of $\langle \mathrm{M1_{LOFAR}} \rangle$, is calculated from the LOFAR data points using bootstrap statistics  \citep{Efron79}. We drew thousands of samples of eight values from the data points. Each value in a sample is one of the data points drawn at random, and a given sample may contain the same data point several times. We compute the mean value for each sample. The PDF that we obtain accounts for the data scatter and the sample variance.

In Fig.~\ref{fig:Faraday_mean_histo}, the PDFs of the mean RM and $\mathrm{M1_{LOFAR}}$ are compared.
We discuss this graph in the reference framework introduced by \citet{Burn:1966} and consider two idealized cases: (1) a Faraday foreground screen of WNM electrons that rotates the polarization angle of a background synchrotron emission, and  (2) a Faraday slab where the polarized synchrotron emission is distributed along the LoS together with the WNM electrons.
In the first case, we expect the mean values of $\mathrm{RM_{WNM}}$ and $\mathrm{M1_{LOFAR}}$ to be the same. In the second case, for a symmetric distribution of thermal electrons and polarized synchrotron emission between the near and far sides of the slab, we expect the mean of $\mathrm{RM_{WNM}}$ to be twice that of $\mathrm{M1_{LOFAR}}$ \citep{Sokoloff98}. 

We find a good match between $\langle \mathrm{RM_{WNM}} \times 0.5 \rangle$ and $\langle \mathrm{M1_{LOFAR}} \rangle$, which favors case (2).  
Individual data points for each star are compared and plotted versus distance in Fig.~\ref{fig:RMstar_vs_distance}. The large error bars on the data points prevent a detailed assessment. We only point out that this plot does not show a systematic trend with distance. The data comparison does not include the systematic uncertainty on electron column densities associated with our choice for $P^\prime_{\rm Ar}$  in Sect.~\ref{subsec:WNM}. We note that within the ionization models quantified by \citet{Jenkins13} the alternative choice for $P^\prime_{\rm Ar}$ would produce a mean difference between the WNM RMs and the first moment of the LOFAR spectra greater than a factor of 2. 

\begin{figure}[!h]
\centering
\includegraphics[scale=0.43]{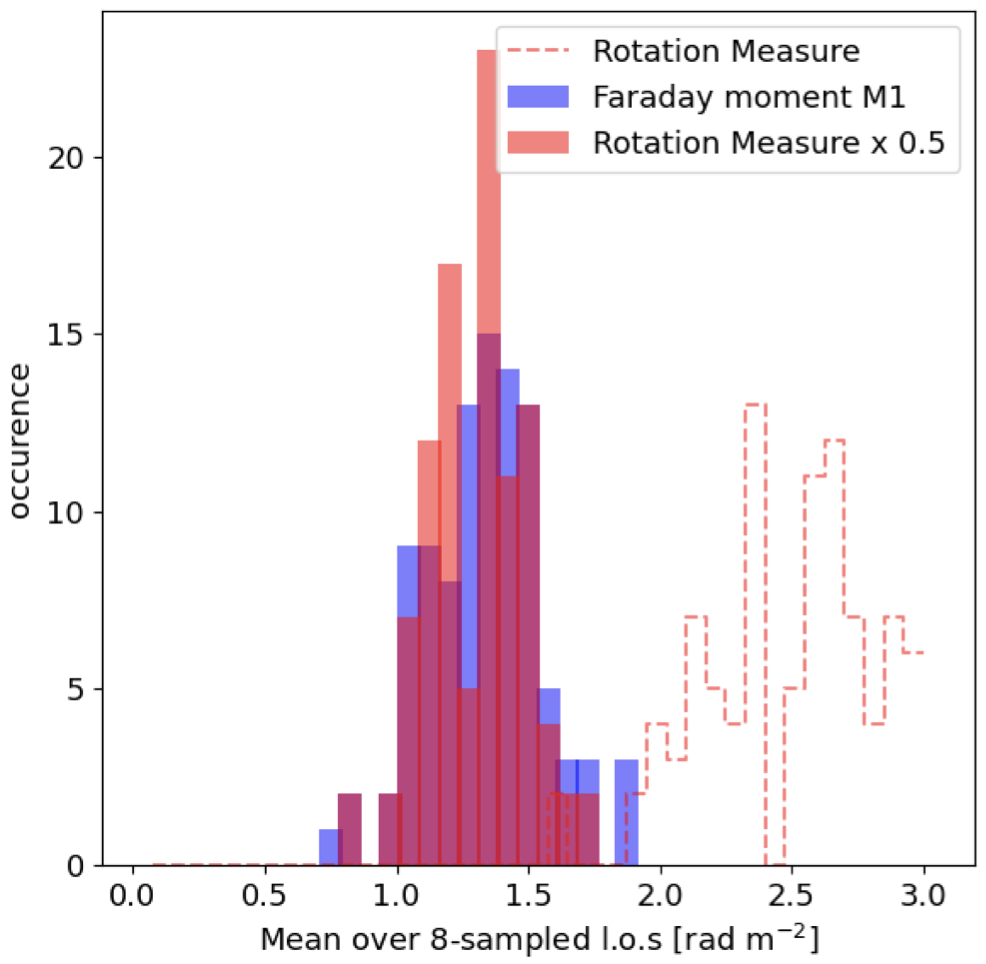}
\caption{PDFs of the
mean RM and the mean LOFAR M1 moments for synthetic observations derived from a numerical simulation by \citet{Bracco:2022}. The mean values are computed on random sets of eight LoS. The PDFs of $\langle \mathrm{RM_{WNM}} \times 0.5 \rangle$ and $\langle M1_\mathrm{LOFAR} \rangle$ are plotted in red and blue. The PDF of $\langle \mathrm{RM_{WNM}} \rangle$ is also shown as a dashed line.}
\label{fig:Faraday_pdfs_sims}
\end{figure}

\begin{table}
%\centering
{\small
    \caption{Data for pulsars within the LoTSS field}
\label{tab:pulsar}
   \begin{tabular}{lccccc}
    \hline
    \hline
    \\[-1.0ex]
Pulsar &  $l $ & $b$ & DM & RM  \\
        $^\circ$  & $^\circ$   & cm$^{-3}$\, pc   & rad\,m$^{-2}$ \\
         &  (a) & (a)   & (b)  & (c) \\       
    \\[-1.0ex]
    \hline \\[-1.0ex]
B0809+74 & 139.998  & 31.618    &  5.75    &  -14.00     \\
J0854+5449 & 162.782 & 39.408   & 18.84       &-8.8     \\
B0917+63  & 151.431  & 40.725   & 13.15    &-14.93     \\
J1012+5307 & 160.347  & 50.858   & 9.02     & 2.98    \\
B1112+50 & 154.408  & 60.365   & 9.18 &   2.41    \\
J1434+7257  & 113.082   & 42.153    & 12.60  &  -9.7   \\
B1508+55  & 91.325   & 52.287    & 19.62   &  1.5 \\
J1518+4904  & 80.808   & 54.282  & 11.61  & -11.9 \\
J1544+4937  &79.172   & 50.166    & 23.23  & 9.8        \\
     \\[-1.0ex]
    \hline
\end{tabular} 
\\[1.0ex]}
(a): Galactic longitudes and latitudes. \\
(b): Dispersion measures. \\
(c): RMs. 
\end{table}

\section{Insight into the interpretation of LOFAR Faraday data}
\label{sec:insight}

Our work contributes to a new perspective on the interpretation of LOFAR Galactic polarization data. We discuss two specific questions: the origin of the LOFAR Faraday structures in Sect.~\ref{subsec:canals} and the contribution of the WIM in Sect.~\ref{subsec:WIM}.

\subsection{Origin of LOFAR Faraday structures }
\label{subsec:canals}

Comparison of the UV and LOFAR data in Fig.~\ref{fig:Faraday_mean_histo} suggests that the relativistic electrons producing the synchrotron-polarized emission detected by LOFAR and the thermal electrons in the WNM contributing to Faraday rotation are mixed within the same ISM slab. We take the median value of $\langle \mathrm{RM_{WNM}} \rangle$ as the mean RM across this slab, $\mathrm{RM}_0 \sim 0.7\, \mathrm{rad\,m}^{-2}$. Because $\mathrm{RM}_0$ is smaller than the resolution in Faraday depth of the LoTSS data ($1.2 \, \mathrm{rad\,m}^{-2}$), the synchrotron emission from the near and far sides of the slab are not separated through the LOFAR RM-synthesis.  

In this context, it is relevant to mention the spatial correlation between polarized dust emission measured by \Planck at $353\,$ GHz and the polarized synchrotron emission measured by \Planck\ and {\it WMAP} at 30 and $23\,$GHz \citep{PIPXXII,Choi15,PlanckXI}. The correlation coefficient, measured to be approximately 20\%, provides an estimate of the fraction of polarized synchrotron emission that comes from the same ISM slab as the WNM, given the tight correlation between dust polarization and \HI\ at high Galactic latitudes \citep{Ghosh17,Clark19}. The remaining 80\% of the polarized synchrotron emission would be background to the WNM electrons. We hypothesize that the background emission at LOFAR frequencies undergoes depolarization due to differential Faraday rotation along the LoS. This idea was previously proposed by \citet{vanEck:2017} and examined in a separate context by \citet{Hill18}. If this view holds, deep LOFAR observations should reveal weak diffuse polarized emission at higher absolute Faraday depths than the $M1_\mathrm{LOFAR}$ values reported by \citet{Erceg22}. 
Analysis of the LoTSS ELAIS-N1 Deep Field at high Galactic latitude, which combines 150 hours of LOFAR observations, does support this hypothesis. \citet{Snidaric23} successfully detected faint Galactic polarized emission at higher absolute Faraday depths, probing the full range of RMs expected from the Galaxy.

 Faraday rotation and depolarization produced by a slab of magnetized ISM have been studied analytically by \citet{Sokoloff98}. The polarized emission depends on the relative distribution of the thermal electrons that produce the Faraday rotation and the relativistic electrons that produce the synchrotron emission. Together with the differential Faraday rotation along the LoS, these factors contribute to the variations of the observed polarized intensity.  However, this formal framework ignores the bandwidth of the LOFAR observations.

To put the data comparison presented in Fig.~\ref{fig:Faraday_mean_histo} into an astrophysical context, we confront our results with those obtained from MHD simulations. To this end, we perform the same comparison between WNM RMs and first moments of Faraday spectra using synthetic observations built from numerical simulations of two colliding shells as input \citep{Ntormousi17}. We use the synthesized LOFAR maps produced and analyzed by \citet{Bracco:2022}. Specifically,  we use their Case~A data where the shells collide in a direction perpendicular to the initial magnetic field for their high value of the cosmic-ray ionization rate $\zeta^H = 2.6 \, 10^{-16}\,\mathrm{s}^{-1}$.  In these simulations, most of the Faraday rotation occurs within partially ionized WNM gas. The synthesized LOFAR maps have morphological features similar to those revealed by LoTSS data, including depolarization features. 

In Fig.~\ref{fig:Faraday_pdfs_sims}, we compare the PDFs of the mean RM and the mean first moment, computed on these synthesized LOFAR maps, for a large number of random sets of eight LoS.  The similarity to the corresponding plot in Fig.~\ref{fig:Faraday_mean_histo} is striking because the simulations were not designed to model our observational work. Additional work is required to investigate the link between Faraday structures and the dynamics of the magnetized multiphase ISM. In particular, it would be interesting to take into account our specific view point within the LB as done for the dust polarization by \citet{Maconi23}, as well as the impact of the supernovae at the origin of the LB \citep{Fuchs06} on the ionization of the local WNM gas \citep{Avillez20} and the associated synchrotron emission.

\subsection{Contribution from the warm ionized medium }
\label{subsec:WIM}

We broaden the context of our work by comparing the column density of electrons in the WNM with that in the WIM. To this purpose, we collected dispersion and rotation measures of pulsars located within the LoTSS field, from the pulsar catalog presented by \citet{Manchester05}\footnote{\url{https://www.atnf.csiro.au/research/pulsar/psrcat/}}. The data are listed in Table~\ref{tab:pulsar}. The catalog does not provide error bars, but those are specified for several of these pulsars by \citet{Sobey19}. These error bars are not significant for our analysis because they are much smaller than the scatter between individual values. 

The mean column density of electrons computed from the dispersion measures is $\langle N_e^\mathrm{Pulsars} \rangle =  4.2 \pm 0.6 \times 10^{19}$cm$^{-2}$, where the error bar is estimated using the bootstrap method \citep{Efron79}. $\langle N_e^\mathrm{Pulsars} \rangle $ is much larger than
$\langle N_e^\mathrm{WNM} \rangle = 5\pm^5_2\times 10^{18}\, \mathrm{cm}^{-2}$ (Sect.~\ref{subsec:WNM}). 
The difference highlights the contribution of the WIM to pulsar dispersion measurements. 
It indicates that the total column density of electrons within the LoTSS field is dominated by the contribution from the WIM. 

Pulsar RMs that range from -15 to $10\, \mathrm{rad\,m}^{-2}$  (Table~\ref{tab:pulsar}) have a wider spread than the LOFAR first moments in the LoTSS field. This observational fact suggests that the WIM is the main source of the total RM of the Galaxy within the LoTSS fields, although this contribution is not easily detectable at LOFAR frequencies. \citet{Erceg22} compared the first LOFAR Faraday moment with those reported by \citet{Dickey19} using data at higher frequencies (1270–1750\,MHz) from the Dominion Radio Astrophysical Observatory (DRAO) and the Galactic RM sky map, derived from observations of polarized extragalactic sources \citep{Hutschenreuter22}. They find that the range of values for the LoTSS first Faraday moment is much smaller than that reported for the DRAO data and the total Galactic RM (see their Fig.~14). 

To account for these results within our astrophysical framework, we propose that the WNM electrons are foreground to the bulk of the WIM electrons. To show that this condition is met, we estimate the distance $L_\mathrm{WNM}$ over which the WNM electrons are distributed.  Combining the electron column density and ionization fraction derived from our data analysis, we obtained: 

\begin{equation}
    L_\mathrm{WNM} = \langle N_e^\mathrm{WNM} \rangle ~ \left(n_H^\mathrm{WNM} \, \langle x_{\rm H}\rangle \, f_\mathrm{WNM} \right)^{-1}, 
\label{eq:ell_WNM_1}
\end{equation}
where $n_H^\mathrm{WNM}$ and  $f_\mathrm{WNM}$ are the Hydrogen density and volume filling factor of the WNM.
Using standard values for these two quantities \citep{Wolfire95}, we find:
\begin{equation}
    L_\mathrm{WNM} \simeq 200 \, \left( \frac{0.4\,\mathrm{H\,cm^{-3}}}{n_H^\mathrm{WNM} }\right) \, \left( \frac{0.2}{f_\mathrm{WNM}}\right) \,\mathrm{pc}.  
\label{eq:ell_WNM_2}
\end{equation}
This estimate of  $L_\mathrm{WNM} $ is much smaller than the scale-height of the WIM  $\sim 1.8\,$kpc  \citep{Gaensler08}.   We also note that it is consistent with the distance estimate for LOFAR structures derived from stellar extinction data by \citet{Turic21} in the 3C196 field, which is close to the LoTSS field in the sky. 

The brightness of diffuse H$\alpha$ emission toward the stars could give some guidance on the WIM in the LoTSS field, but these observations combine foreground and background contributions. Appendix~\ref{App_sec:NII} presents an attempt to estimate the emission measure (EM) of ionized gas foreground to the stars, using absorption lines in FUSE spectra of ionized nitrogen in its second excited fine-structure state. The FUSE spectra provide upper limits and two detections for HD113001 and Ton102, which are all significantly higher than the EM of the full LoS derived from the H$\alpha$ emission. Thus, these EM estimates prove to be of very little significance for our study.

\section{Summary}
\label{sec:conclusion}

We have analyzed stellar UV spectroscopic observations that provide us with estimates of electron column densities $N_e^\mathrm{WNM}$ in the WNM foreground to nine stars within the sky area of the LoTSS mosaic presented by \citet{Erceg22}. The stellar distances range from 140 to 1350\,pc. The UV data are combined with a model of the local magnetic field fitted to \Planck\ maps of dust polarization. We obtain estimates of the RM to the stars from electrons in the WNM, which we compare to LOFAR Faraday spectra. We draw the following results from the data analysis and comparison. 

\begin{itemize}

\item
The mean electron column density in the partially ionized WNM foreground to the stars is $ \langle N_e^\mathrm{WNM} \rangle = 5\pm^5_2\times 10^{18}\,\mathrm{cm}^{-2}$. The mean ionization fraction of hydrogen is $\langle x_{\rm H}\rangle = 0.097\pm^{0.133}_{0.037}$. \\

\item
The first moment of the LOFAR spectra at the positions of our stars are on average half the WNM RMs. This is the result expected for a slab of magnetized ISM where the thermal electrons that produce the Faraday rotation and the relativistic electrons that radiate synchrotron emission are mixed with a roughly symmetric distribution along the LoS.  \\

\item 
The mean RM across the WNM slab is $\mathrm{RM}_0 \sim 0.7\, \mathrm{rad\,m}^{-2}$. As $\mathrm{RM}_0$ is smaller than the resolution in Faraday depth of the LoTSS data, the synchrotron emission from the near and far sides of the slab are not separated through the LOFAR RM-synthesis. \\

\item 
We broadened the context of our work by comparing the column density of electrons in the WNM with dispersion measures of pulsars located within the same LoTSS field. The mean electron column density computed from the dispersion measures $\langle N_e^\mathrm{Pulsars} \rangle =  4.2 \pm 0.6 \times 10^{19}$cm$^{-2}$ is about one order of magnitude larger than
$\langle N_e^\mathrm{WNM} \rangle $. This difference highlights the contribution of the WIM to pulsar dispersion measures. \\

\item
Although the WIM constitutes the majority of the total electron column density in the LoTSS field, its impact on the RM at LOFAR frequencies is elusive. To account for this observational fact, we propose that the WNM electrons and the associated polarized synchrotron emission are foreground to the bulk of the WIM. 

\end{itemize}

Our work sheds new light on the interpretation of the LOFAR Galactic polarization data. 
This suggests that LOFAR Faraday structures are associated with neutral, partially ionized, WNM. 
Pending confirmation with a larger sample of stars when the full LOFAR northern sky maps are published, this preliminary finding lays an astrophysics framework for exploring the relationship between Faraday structures and the dynamics of the magnetized multiphase interstellar medium.
Independently of the LOFAR data, the analysis of UV FUSE data for a larger sample of stars over the whole sky can provide valuable information on the distance distribution of WNM electrons in the local ISM.  

\begin{acknowledgements}
This research was supported by the COGITO No. 42816 exchange program between France and Croatia. AB and FB acknowledge the support of the European Research Council, under the Seventh Framework Program of the European Community, through the Advanced Grant MIST (FP7/2017-2022, No. 787813). AB also acknowledges financial support from the INAF initiative "IAF Astronomy Fellowships in Italy" (grant name MEGASKAT). We thank M. Alves, K. Ferri\`ere and P. Lesaffre for helpful discussions, and the anonymous referee whose report helped us significantly improve the original version of the article.  
\end{acknowledgements}

\bibliographystyle{aa}
\bibliography{Refs_UV_LOFAR}

\begin{thebibliography}{78}
\expandafter\ifx\csname natexlab\endcsname\relax\def\natexlab#1{#1}\fi

\bibitem[{{Bailer-Jones} {et~al.}(2021){Bailer-Jones}, {Rybizki}, {Fouesneau},
  {Demleitner}, \& {Andrae}}]{Bailer-Jones21}
{Bailer-Jones}, C.~A.~L., {Rybizki}, J., {Fouesneau}, M., {Demleitner}, M., \&
  {Andrae}, R. 2021, \aj, 161, 147

\bibitem[{{Beck} {et~al.}(2003){Beck}, {Shukurov}, {Sokoloff}, \&
  {Wielebinski}}]{Beck03}
{Beck}, R., {Shukurov}, A., {Sokoloff}, D., \& {Wielebinski}, R. 2003, \aap,
  411, 99

\bibitem[{{Bracco} {et~al.}(2020){Bracco}, {Jeli{\'c}}, {Marchal}, {Turi{\'c}},
  {Erceg}, {Miville-Desch{\^e}nes}, \& {Bellomi}}]{Bracco20}
{Bracco}, A., {Jeli{\'c}}, V., {Marchal}, A., {et~al.} 2020, \aap, 644, L3

\bibitem[{{Bracco} {et~al.}(2022){Bracco}, {Ntormousi}, {Jeli{\'c}},
  {Padovani}, {{\v{S}}iljeg}, {Erceg}, {Turi{\'c}}, {Ceraj}, \&
  {{\v{S}}nidari{\'c}}}]{Bracco:2022}
{Bracco}, A., {Ntormousi}, E., {Jeli{\'c}}, V., {et~al.} 2022, \aap, 663, A37

\bibitem[{{Brentjens} \& {de Bruyn}(2005)}]{Brentjens&deBruyn:2005}
{Brentjens}, M.~A. \& {de Bruyn}, A.~G. 2005, \aap, 441, 1217

\bibitem[{{Burn}(1966)}]{Burn:1966}
{Burn}, B.~J. 1966, \mnras, 133, 67

\bibitem[{{Bzowski} {et~al.}(2019){Bzowski}, {Czechowski}, {Frisch},
  {Fuselier}, {Galli}, {Grygorczuk}, {Heerikhuisen}, {Kubiak}, {Kucharek},
  {McComas}, {M{\"o}bius}, {Schwadron}, {Slavin}, {Sok{\'o}{\l}}, {Swaczyna},
  {Wurz}, \& {Zirnstein}}]{Bzowski19}
{Bzowski}, M., {Czechowski}, A., {Frisch}, P.~C., {et~al.} 2019, \apj, 882, 60

\bibitem[{{Choi} \& {Page}(2015)}]{Choi15}
{Choi}, S.~K. \& {Page}, L.~A. 2015, \jcap, 2015, 020

\bibitem[{{Clark} \& {Hensley}(2019)}]{Clark19}
{Clark}, S.~E. \& {Hensley}, B.~S. 2019, \apj, 887, 136

\bibitem[{{de Avillez} {et~al.}(2020){de Avillez}, {Anela}, {Asgekar},
  {Breitschwerdt}, \& {Schnitzeler}}]{Avillez20}
{de Avillez}, M.~A., {Anela}, G.~J., {Asgekar}, A., {Breitschwerdt}, D., \&
  {Schnitzeler}, D. H.~F.~M. 2020, \aap, 644, A156

\bibitem[{{Delouis} {et~al.}(2019){Delouis}, {Pagano}, {Mottet}, {Puget}, \&
  {Vibert}}]{Delouis19}
{Delouis}, J.~M., {Pagano}, L., {Mottet}, S., {Puget}, J.~L., \& {Vibert}, L.
  2019, \aap, 629, A38

\bibitem[{{Dickey} {et~al.}(2019){Dickey}, {Landecker}, {Thomson}, {Wolleben},
  {Sun}, {Carretti}, {Douglas}, {Fletcher}, {Gaensler}, {Gray}, {Haverkorn},
  {Hill}, {Mao}, \& {McClure-Griffiths}}]{Dickey19}
{Dickey}, J.~M., {Landecker}, T.~L., {Thomson}, A. J.~M., {et~al.} 2019, \apj,
  871, 106

\bibitem[{{Draine}(2011)}]{Draine11}
{Draine}, B.~T. 2011, {Physics of the Interstellar and Intergalactic Medium,
  Chapter 9}

\bibitem[{{Efron}(1979)}]{Efron79}
{Efron}, B. 1979, Ann. Statistics, 7, 1

\bibitem[{{Erceg} {et~al.}(2022){Erceg}, {Jeli{\'c}}, {Haverkorn}, {Bracco},
  {Shimwell}, {Tasse}, {Dickey}, {Ceraj}, {Drabent}, {Hardcastle}, \&
  {Turi{\'c}}}]{Erceg22}
{Erceg}, A., {Jeli{\'c}}, V., {Haverkorn}, M., {et~al.} 2022, \aap, 663, A7

\bibitem[{{Foreman-Mackey} {et~al.}(2013){Foreman-Mackey}, {Hogg}, {Lang}, \&
  {Goodman}}]{EMCEE13}
{Foreman-Mackey}, D., {Hogg}, D.~W., {Lang}, D., \& {Goodman}, J. 2013, \pasp,
  125, 306

\bibitem[{{Fuchs} {et~al.}(2006){Fuchs}, {Breitschwerdt}, {de Avillez},
  {Dettbarn}, \& {Flynn}}]{Fuchs06}
{Fuchs}, B., {Breitschwerdt}, D., {de Avillez}, M.~A., {Dettbarn}, C., \&
  {Flynn}, C. 2006, \mnras, 373, 993

\bibitem[{{Gaensler} {et~al.}(2008){Gaensler}, {Madsen}, {Chatterjee}, \&
  {Mao}}]{Gaensler08}
{Gaensler}, B.~M., {Madsen}, G.~J., {Chatterjee}, S., \& {Mao}, S.~A. 2008,
  \pasa, 25, 184

\bibitem[{{Ghosh} {et~al.}(2017){Ghosh}, {Boulanger}, {Martin}, {Bracco},
  {Vansyngel}, {Aumont}, {Bock}, {Dor{\'e}}, {Haud}, {Kalberla}, \&
  {Serra}}]{Ghosh17}
{Ghosh}, T., {Boulanger}, F., {Martin}, P.~G., {et~al.} 2017, \aap, 601, A71

\bibitem[{{Goldsmith} {et~al.}(2015){Goldsmith}, {Y{\i}ld{\i}z}, {Langer}, \&
  {Pineda}}]{Goldsmith15}
{Goldsmith}, P.~F., {Y{\i}ld{\i}z}, U.~A., {Langer}, W.~D., \& {Pineda}, J.~L.
  2015, \apj, 814, 133

\bibitem[{{G{\'o}rski} {et~al.}(2005){G{\'o}rski}, {Hivon}, {Banday},
  {Wandelt}, {Hansen}, {Reinecke}, \& {Bartelmann}}]{Gorski05}
{G{\'o}rski}, K.~M., {Hivon}, E., {Banday}, A.~J., {et~al.} 2005, \apj, 622,
  759

\bibitem[{{Gry} \& {Jenkins}(2017)}]{Gry17}
{Gry}, C. \& {Jenkins}, E.~B. 2017, \aap, 598, A31

\bibitem[{{Haffner} {et~al.}(2003){Haffner}, {Reynolds}, {Tufte}, {Madsen},
  {Jaehnig}, \& {Percival}}]{Haffner03}
{Haffner}, L.~M., {Reynolds}, R.~J., {Tufte}, S.~L., {et~al.} 2003, \apjs, 149,
  405

\bibitem[{{Heiles} \& {Haverkorn}(2012)}]{Heiles_Haverkorn12}
{Heiles}, C. \& {Haverkorn}, M. 2012, \ssr, 166, 293

\bibitem[{{HI4PI Collaboration} {et~al.}(2016){HI4PI Collaboration}, {Ben
  Bekhti}, {Fl{\"o}er}, {Keller}, {Kerp}, {Lenz}, {Winkel}, {Bailin},
  {Calabretta}, {Dedes}, {Ford}, {Gibson}, {Haud}, {Janowiecki}, {Kalberla},
  {Lockman}, {McClure-Griffiths}, {Murphy}, {Nakanishi}, {Pisano}, \&
  {Staveley-Smith}}]{HI4Pi16}
{HI4PI Collaboration}, {Ben Bekhti}, N., {Fl{\"o}er}, L., {et~al.} 2016, \aap,
  594, A116

\bibitem[{{Hill}(2018)}]{Hill18}
{Hill}, A. 2018, Galaxies, 6, 129

\bibitem[{{Hutschenreuter} {et~al.}(2022){Hutschenreuter}, {Anderson}, {Betti},
  {Bower}, {Brown}, {Br{\"u}ggen}, {Carretti}, {Clarke}, {Clegg}, {Costa},
  {Croft}, {Van Eck}, {Gaensler}, {de Gasperin}, {Haverkorn}, {Heald}, {Hull},
  {Inoue}, {Johnston-Hollitt}, {Kaczmarek}, {Law}, {Ma}, {MacMahon}, {Mao},
  {Riseley}, {Roy}, {Shanahan}, {Shimwell}, {Stil}, {Sobey}, {O'Sullivan},
  {Tasse}, {Vacca}, {Vernstrom}, {Williams}, {Wright}, \&
  {En{\ss}lin}}]{Hutschenreuter22}
{Hutschenreuter}, S., {Anderson}, C.~S., {Betti}, S., {et~al.} 2022, \aap, 657,
  A43

\bibitem[{{Iacobelli} {et~al.}(2013){Iacobelli}, {Haverkorn}, \&
  {Katgert}}]{Iacobelli13}
{Iacobelli}, M., {Haverkorn}, M., \& {Katgert}, P. 2013, \aap, 549, A56

\bibitem[{{Jeli{\'c}} {et~al.}(2014){Jeli{\'c}}, {de Bruyn}, {Mevius},
  {Abdalla}, {Asad}, {Bernardi}, {Brentjens}, {Bus}, {Chapman}, {Ciardi},
  {Daiboo}, {Fernandez}, {Ghosh}, {Harker}, {Jensen}, {Kazemi}, {Koopmans},
  {Labropoulos}, {Martinez-Rubi}, {Mellema}, {Offringa}, {Pandey}, {Patil},
  {Thomas}, {Vedantham}, {Veligatla}, {Yatawatta}, {Zaroubi}, {Alexov},
  {Anderson}, {Avruch}, {Beck}, {Bell}, {Bentum}, {Best}, {Bonafede},
  {Bregman}, {Breitling}, {Broderick}, {Brouw}, {Br{\"u}ggen}, {Butcher},
  {Conway}, {de Gasperin}, {de Geus}, {Deller}, {Dettmar}, {Duscha},
  {Eisl{\"o}ffel}, {Engels}, {Falcke}, {Fallows}, {Fender}, {Ferrari},
  {Frieswijk}, {Garrett}, {Grie{\ss}meier}, {Gunst}, {Hamaker}, {Hassall},
  {Haverkorn}, {Heald}, {Hessels}, {Hoeft}, {H{\"o}randel}, {Horneffer}, {van
  der Horst}, {Iacobelli}, {Juette}, {Karastergiou}, {Kondratiev}, {Kramer},
  {Kuniyoshi}, {Kuper}, {van Leeuwen}, {Maat}, {Mann}, {McKay-Bukowski},
  {McKean}, {Munk}, {Nelles}, {Norden}, {Paas}, {Pandey-Pommier}, {Pietka},
  {Pizzo}, {Polatidis}, {Reich}, {R{\"o}ttgering}, {Rowlinson}, {Scaife},
  {Schwarz}, {Serylak}, {Smirnov}, {Steinmetz}, {Stewart}, {Tagger}, {Tang},
  {Tasse}, {ter Veen}, {Thoudam}, {Toribio}, {Vermeulen}, {Vocks}, {van
  Weeren}, {Wijers}, {Wijnholds}, {Wucknitz}, \& {Zarka}}]{Jelic:2014}
{Jeli{\'c}}, V., {de Bruyn}, A.~G., {Mevius}, M., {et~al.} 2014, \aap, 568,
  A101

\bibitem[{{Jeli{\'c}} {et~al.}(2015){Jeli{\'c}}, {de Bruyn}, {Pandey},
  {Mevius}, {Haverkorn}, {Brentjens}, {Koopmans}, {Zaroubi}, {Abdalla}, {Asad},
  {Bus}, {Chapman}, {Ciardi}, {Fernandez}, {Ghosh}, {Harker}, {Iliev},
  {Jensen}, {Kazemi}, {Mellema}, {Offringa}, {Patil}, {Vedantham}, \&
  {Yatawatta}}]{Jelic:2015}
{Jeli{\'c}}, V., {de Bruyn}, A.~G., {Pandey}, V.~N., {et~al.} 2015, \aap, 583,
  A137

\bibitem[{{Jeli{\'c}} {et~al.}(2018){Jeli{\'c}}, {Prelogovi{\'c}}, {Haverkorn},
  {Remeijn}, \& {Klind{\v{z}}i{\'c}}}]{Jelic:2018}
{Jeli{\'c}}, V., {Prelogovi{\'c}}, D., {Haverkorn}, M., {Remeijn}, J., \&
  {Klind{\v{z}}i{\'c}}, D. 2018, \aap, 615, L3

\bibitem[{{Jenkins}(2009)}]{Jenkins09}
{Jenkins}, E.~B. 2009, \apj, 700, 1299

\bibitem[{{Jenkins}(2013)}]{Jenkins13}
{Jenkins}, E.~B. 2013, \apj, 764, 25

\bibitem[{{Kalberla} \& {Haud}(2018)}]{Kalberla18}
{Kalberla}, P.~M.~W. \& {Haud}, U. 2018, \aap, 619, A58

\bibitem[{{Kalberla} \& {Kerp}(2016)}]{Kalberla16}
{Kalberla}, P.~M.~W. \& {Kerp}, J. 2016, \aap, 595, A37

\bibitem[{{Kalberla} {et~al.}(2017){Kalberla}, {Kerp}, {Haud}, \&
  {Haverkorn}}]{Kalberla:2017}
{Kalberla}, P.~M.~W., {Kerp}, J., {Haud}, U., \& {Haverkorn}, M. 2017, \aap,
  607, A15

\bibitem[{{Kruk} {et~al.}(2002){Kruk}, {Howk}, {Andr{\'e}}, {Moos}, {Oegerle},
  {Oliveira}, {Sembach}, {Chayer}, {Linsky}, {Wood}, {Ferlet}, {H{\'e}brard},
  {Lemoine}, {Vidal-Madjar}, \& {Sonneborn}}]{Kruk02}
{Kruk}, J.~W., {Howk}, J.~C., {Andr{\'e}}, M., {et~al.} 2002, ApJ Suppl, 140,
  19

\bibitem[{{Lallement} {et~al.}(2014){Lallement}, {Vergely}, {Valette},
  {Puspitarini}, {Eyer}, \& {Casagrande}}]{Lallement14}
{Lallement}, R., {Vergely}, J.~L., {Valette}, B., {et~al.} 2014, \aap, 561, A91

\bibitem[{{Lenc} {et~al.}(2016){Lenc}, {Gaensler}, {Sun}, {Sadler}, {Willis},
  {Barry}, {Beardsley}, {Bell}, {Bernardi}, {Bowman}, {Briggs}, {Callingham},
  {Cappallo}, {Carroll}, {Corey}, {de Oliveira-Costa}, {Deshpande}, {Dillon},
  {Dwarkanath}, {Emrich}, {Ewall-Wice}, {Feng}, {For}, {Goeke}, {Greenhill},
  {Hancock}, {Hazelton}, {Hewitt}, {Hindson}, {Hurley-Walker},
  {Johnston-Hollitt}, {Jacobs}, {Kapi{\'n}ska}, {Kaplan}, {Kasper}, {Kim},
  {Kratzenberg}, {Line}, {Loeb}, {Lonsdale}, {Lynch}, {McKinley}, {McWhirter},
  {Mitchell}, {Morales}, {Morgan}, {Morgan}, {Murphy}, {Neben}, {Oberoi},
  {Offringa}, {Ord}, {Paul}, {Pindor}, {Pober}, {Prabu}, {Procopio}, {Riding},
  {Rogers}, {Roshi}, {Udaya Shankar}, {Sethi}, {Srivani}, {Staveley-Smith},
  {Subrahmanyan}, {Sullivan}, {Tegmark}, {Thyagarajan}, {Tingay}, {Trott},
  {Waterson}, {Wayth}, {Webster}, {Whitney}, {Williams}, {Williams}, {Wu},
  {Wyithe}, \& {Zheng}}]{Lenc16}
{Lenc}, E., {Gaensler}, B.~M., {Sun}, X.~H., {et~al.} 2016, \apj, 830, 38

\bibitem[{{Lindegren} {et~al.}(2021){Lindegren}, {Klioner}, {Hern{\'a}ndez},
  {Bombrun}, {Ramos-Lerate}, {Steidelm{\"u}ller}, {Bastian}, {Biermann}, {de
  Torres}, {Gerlach}, {Geyer}, {Hilger}, {Hobbs}, {Lammers}, {McMillan},
  {Stephenson}, {Casta{\~n}eda}, {Davidson}, {Fabricius}, {Gracia-Abril},
  {Portell}, {Rowell}, {Teyssier}, {Torra}, {Bartolom{\'e}}, {Clotet},
  {Garralda}, {Gonz{\'a}lez-Vidal}, {Torra}, {Abbas}, {Altmann}, {Anglada
  Varela}, {Balaguer-N{\'u}{\~n}ez}, {Balog}, {Barache}, {Becciani}, {Bernet},
  {Bertone}, {Bianchi}, {Bouquillon}, {Brown}, {Bucciarelli}, {Busonero},
  {Butkevich}, {Buzzi}, {Cancelliere}, {Carlucci}, {Charlot}, {Cioni},
  {Crosta}, {Crowley}, {del Peloso}, {del Pozo}, {Drimmel}, {Esquej}, {Fienga},
  {Fraile}, {Gai}, {Garcia-Reinaldos}, {Guerra}, {Hambly}, {Hauser},
  {Jan{\ss}en}, {Jordan}, {Kostrzewa-Rutkowska}, {Lattanzi}, {Liao}, {Licata},
  {Lister}, {L{\"o}ffler}, {Marchant}, {Masip}, {Mignard}, {Mints}, {Molina},
  {Mora}, {Morbidelli}, {Murphy}, {Pagani}, {Panuzzo}, {Pe{\~n}alosa Esteller},
  {Poggio}, {Re Fiorentin}, {Riva}, {Sagrist{\`a} Sell{\'e}s}, {Sanchez
  Gimenez}, {Sarasso}, {Sciacca}, {Siddiqui}, {Smart}, {Souami}, {Spagna},
  {Steele}, {Taris}, {Utrilla}, {van Reeven}, \& {Vecchiato}}]{Lindegren21}
{Lindegren}, L., {Klioner}, S.~A., {Hern{\'a}ndez}, J., {et~al.} 2021, \aap,
  649, A2

\bibitem[{{Maconi} {et~al.}(2023){Maconi}, {Soler}, {Reissl}, {Girichidis},
  {Klessen}, {Hennebelle}, {Molinari}, {Testi}, {Smith}, {Sormani}, {Teh}, \&
  {Traficante}}]{Maconi23}
{Maconi}, E., {Soler}, J.~D., {Reissl}, S., {et~al.} 2023, \mnras, 523, 5995

\bibitem[{{Manchester} {et~al.}(2005){Manchester}, {Hobbs}, {Teoh}, \&
  {Hobbs}}]{Manchester05}
{Manchester}, R.~N., {Hobbs}, G.~B., {Teoh}, A., \& {Hobbs}, M. 2005, \aj, 129,
  1993

\bibitem[{{Morton}(2003)}]{Morton03}
{Morton}, D.~C. 2003, \apjs, 149, 205

\bibitem[{{Ntormousi} {et~al.}(2017){Ntormousi}, {Dawson}, {Hennebelle}, \&
  {Fierlinger}}]{Ntormousi17}
{Ntormousi}, E., {Dawson}, J.~R., {Hennebelle}, P., \& {Fierlinger}, K. 2017,
  \aap, 599, A94

\bibitem[{{Planck 2018 results III}(2020)}]{PlanckIII}
{Planck 2018 results III}. 2020, \aap, 641, A3

\bibitem[{{Planck 2018 results XI}(2020)}]{PlanckXI}
{Planck 2018 results XI}. 2020, \aap, 641, A11

\bibitem[{{Planck 2018 Results XII}(2020)}]{PlanckXII}
{Planck 2018 Results XII}. 2020, \aap, 641, A12

\bibitem[{{Planck Intermediate Results XLIV}(2016)}]{PIPXLIV}
{Planck Intermediate Results XLIV}. 2016, \aap, 596, A105

\bibitem[{{Planck Intermediate Results XXII}(2015)}]{PIPXXII}
{Planck Intermediate Results XXII}. 2015, \aap, 576, A107

\bibitem[{{Redfield} \& {Falcon}(2008)}]{Redfield08b}
{Redfield}, S. \& {Falcon}, R.~E. 2008, \apj, 683, 207

\bibitem[{{Regaldo-Saint Blancard} {et~al.}(2020){Regaldo-Saint Blancard},
  {Levrier}, {Allys}, {Bellomi}, \& {Boulanger}}]{Regaldo20}
{Regaldo-Saint Blancard}, B., {Levrier}, F., {Allys}, E., {Bellomi}, E., \&
  {Boulanger}, F. 2020, \aap, 642, A217

\bibitem[{{Reynolds} {et~al.}(1998){Reynolds}, {Tufte}, {Haffner}, {Jaehnig},
  \& {Percival}}]{Reynolds:1998}
{Reynolds}, R.~J., {Tufte}, S.~L., {Haffner}, L.~M., {Jaehnig}, K., \&
  {Percival}, J.~W. 1998, \pasa, 15, 14

\bibitem[{{Shimwell} {et~al.}(2022){Shimwell}, {Hardcastle}, {Tasse}, {Best},
  {R{\"o}ttgering}, {Williams}, {Botteon}, {Drabent}, {Mechev}, {Shulevski},
  {van Weeren}, {Bester}, {Br{\"u}ggen}, {Brunetti}, {Callingham}, {Chy{\.z}y},
  {Conway}, {Dijkema}, {Duncan}, {de Gasperin}, {Hale}, {Haverkorn}, {Hugo},
  {Jackson}, {Mevius}, {Miley}, {Morabito}, {Morganti}, {Offringa}, {Oonk},
  {Rafferty}, {Sabater}, {Smith}, {Schwarz}, {Smirnov}, {O'Sullivan},
  {Vedantham}, {White}, {Albert}, {Alegre}, {Asabere}, {Bacon}, {Bonafede},
  {Bonnassieux}, {Brienza}, {Bilicki}, {Bonato}, {Calistro Rivera}, {Cassano},
  {Cochrane}, {Croston}, {Cuciti}, {Dallacasa}, {Danezi}, {Dettmar}, {Di
  Gennaro}, {Edler}, {En{\ss}lin}, {Emig}, {Franzen}, {Garc{\'\i}a-Vergara},
  {Grange}, {G{\"u}rkan}, {Hajduk}, {Heald}, {Heesen}, {Hoang}, {Hoeft},
  {Horellou}, {Iacobelli}, {Jamrozy}, {Jeli{\'c}}, {Kondapally}, {Kukreti},
  {Kunert-Bajraszewska}, {Magliocchetti}, {Mahatma}, {Ma{\l}ek}, {Mandal},
  {Massaro}, {Meyer-Zhao}, {Mingo}, {Mostert}, {Nair}, {Nakoneczny},
  {Nikiel-Wroczy{\'n}ski}, {Orr{\'u}}, {Pajdosz-{\'S}mierciak}, {Pasini},
  {Prandoni}, {van Piggelen}, {Rajpurohit}, {Retana-Montenegro}, {Riseley},
  {Rowlinson}, {Saxena}, {Schrijvers}, {Sweijen}, {Siewert}, {Timmerman},
  {Vaccari}, {Vink}, {West}, {Wo{\l}owska}, {Zhang}, \& {Zheng}}]{Shimwell22}
{Shimwell}, T.~W., {Hardcastle}, M.~J., {Tasse}, C., {et~al.} 2022, \aap, 659,
  A1

\bibitem[{{Shimwell} {et~al.}(2017){Shimwell}, {R{\"o}ttgering}, {Best},
  {Williams}, {Dijkema}, {de Gasperin}, {Hardcastle}, {Heald}, {Hoang},
  {Horneffer}, {Intema}, {Mahony}, {Mandal}, {Mechev}, {Morabito}, {Oonk},
  {Rafferty}, {Retana-Montenegro}, {Sabater}, {Tasse}, {van Weeren},
  {Br{\"u}ggen}, {Brunetti}, {Chy{\.z}y}, {Conway}, {Haverkorn}, {Jackson},
  {Jarvis}, {McKean}, {Miley}, {Morganti}, {White}, {Wise}, {van Bemmel},
  {Beck}, {Brienza}, {Bonafede}, {Calistro Rivera}, {Cassano}, {Clarke},
  {Cseh}, {Deller}, {Drabent}, {van Driel}, {Engels}, {Falcke}, {Ferrari},
  {Fr{\"o}hlich}, {Garrett}, {Harwood}, {Heesen}, {Hoeft}, {Horellou},
  {Israel}, {Kapi{\'n}ska}, {Kunert-Bajraszewska}, {McKay}, {Mohan},
  {Orr{\'u}}, {Pizzo}, {Prandoni}, {Schwarz}, {Shulevski}, {Sipior}, {Smith},
  {Sridhar}, {Steinmetz}, {Stroe}, {Varenius}, {van der Werf}, {Zensus}, \&
  {Zwart}}]{Shimwell17}
{Shimwell}, T.~W., {R{\"o}ttgering}, H.~J.~A., {Best}, P.~N., {et~al.} 2017,
  \aap, 598, A104

\bibitem[{{Shimwell} {et~al.}(2019){Shimwell}, {Tasse}, {Hardcastle}, {Mechev},
  {Williams}, {Best}, {R{\"o}ttgering}, {Callingham}, {Dijkema}, {de Gasperin},
  {Hoang}, {Hugo}, {Mirmont}, {Oonk}, {Prandoni}, {Rafferty}, {Sabater},
  {Smirnov}, {van Weeren}, {White}, {Atemkeng}, {Bester}, {Bonnassieux},
  {Br{\"u}ggen}, {Brunetti}, {Chy{\.z}y}, {Cochrane}, {Conway}, {Croston},
  {Danezi}, {Duncan}, {Haverkorn}, {Heald}, {Iacobelli}, {Intema}, {Jackson},
  {Jamrozy}, {Jarvis}, {Lakhoo}, {Mevius}, {Miley}, {Morabito}, {Morganti},
  {Nisbet}, {Orr{\'u}}, {Perkins}, {Pizzo}, {Schrijvers}, {Smith}, {Vermeulen},
  {Wise}, {Alegre}, {Bacon}, {van Bemmel}, {Beswick}, {Bonafede}, {Botteon},
  {Bourke}, {Brienza}, {Calistro Rivera}, {Cassano}, {Clarke}, {Conselice},
  {Dettmar}, {Drabent}, {Dumba}, {Emig}, {En{\ss}lin}, {Ferrari}, {Garrett},
  {G{\'e}nova-Santos}, {Goyal}, {G{\"u}rkan}, {Hale}, {Harwood}, {Heesen},
  {Hoeft}, {Horellou}, {Jackson}, {Kokotanekov}, {Kondapally},
  {Kunert-Bajraszewska}, {Mahatma}, {Mahony}, {Mandal}, {McKean}, {Merloni},
  {Mingo}, {Miskolczi}, {Mooney}, {Nikiel-Wroczy{\'n}ski}, {O'Sullivan},
  {Quinn}, {Reich}, {Roskowi{\'n}ski}, {Rowlinson}, {Savini}, {Saxena},
  {Schwarz}, {Shulevski}, {Sridhar}, {Stacey}, {Urquhart}, {van der Wiel},
  {Varenius}, {Webster}, \& {Wilber}}]{Shimwell19}
{Shimwell}, T.~W., {Tasse}, C., {Hardcastle}, M.~J., {et~al.} 2019, \aap, 622,
  A1

\bibitem[{{Skalidis} \& {Pelgrims}(2019)}]{Skalidis19}
{Skalidis}, R. \& {Pelgrims}, V. 2019, \aap, 631, L11

\bibitem[{{Slavin} \& {Frisch}(2008)}]{Slavin08}
{Slavin}, J.~D. \& {Frisch}, P.~C. 2008, \aap, 491, 53

\bibitem[{{Slavin} {et~al.}(2000){Slavin}, {McKee}, \& {Hollenbach}}]{Slavin00}
{Slavin}, J.~D., {McKee}, C.~F., \& {Hollenbach}, D.~J. 2000, \apj, 541, 218

\bibitem[{{Sobey} {et~al.}(2019){Sobey}, {Bilous}, {Grie{\ss}meier}, {Hessels},
  {Karastergiou}, {Keane}, {Kondratiev}, {Kramer}, {Michilli}, {Noutsos},
  {Pilia}, {Polzin}, {Stappers}, {Tan}, {van Leeuwen}, {Verbiest},
  {Weltevrede}, {Heald}, {Alves}, {Carretti}, {En{\ss}lin}, {Haverkorn},
  {Iacobelli}, {Reich}, \& {Van Eck}}]{Sobey19}
{Sobey}, C., {Bilous}, A.~V., {Grie{\ss}meier}, J.~M., {et~al.} 2019, \mnras,
  484, 3646

\bibitem[{{Sofia} \& {Jenkins}(1998)}]{Sofia98}
{Sofia}, U.~J. \& {Jenkins}, E.~B. 1998, \apj, 499, 951

\bibitem[{{Sokoloff} {et~al.}(1998){Sokoloff}, {Bykov}, {Shukurov},
  {Berkhuijsen}, {Beck}, \& {Poezd}}]{Sokoloff98}
{Sokoloff}, D.~D., {Bykov}, A.~A., {Shukurov}, A., {et~al.} 1998, \mnras, 299,
  189

\bibitem[{{Spitzer} \& {Fitzpatrick}(1993)}]{Spitzer93}
{Spitzer}, Lyman, J. \& {Fitzpatrick}, E.~L. 1993, \apj, 409, 299

\bibitem[{{Str{\"o}mgren}(1948)}]{Stromgren48}
{Str{\"o}mgren}, B. 1948, \apj, 108, 242

\bibitem[{{Sutherland} \& {Dopita}(2017)}]{Sutherland:2017}
{Sutherland}, R.~S. \& {Dopita}, M.~A. 2017, \apjs, 229, 34

\bibitem[{{Tayal}(2011)}]{Tayal11}
{Tayal}, S.~S. 2011, \apjs, 195, 12

\bibitem[{{Thomson} {et~al.}(2019){Thomson}, {Landecker}, {Dickey},
  {McClure-Griffiths}, {Wolleben}, {Carretti}, {Fletcher}, {Federrath}, {Hill},
  {Mao}, {Gaensler}, {Haverkorn}, {Clark}, {Van Eck}, \& {West}}]{Thomson19}
{Thomson}, A. J.~M., {Landecker}, T.~L., {Dickey}, J.~M., {et~al.} 2019,
  \mnras, 487, 4751

\bibitem[{{Turi{\'c}} {et~al.}(2021){Turi{\'c}}, {Jeli{\'c}}, {Jaspers},
  {Haverkorn}, {Bracco}, {Erceg}, {Ceraj}, {van Eck}, \& {Zaroubi}}]{Turic21}
{Turi{\'c}}, L., {Jeli{\'c}}, V., {Jaspers}, R., {et~al.} 2021, \aap, 654, A5

\bibitem[{{Van Eck}(2018)}]{vanEck18}
{Van Eck}, C. 2018, Galaxies, 6, 112

\bibitem[{{Van Eck} {et~al.}(2019){Van Eck}, {Haverkorn}, {Alves}, {Beck},
  {Best}, {Carretti}, {Chy{\.z}y}, {En{\ss}lin}, {Farnes}, {Ferri{\`e}re},
  {Heald}, {Iacobelli}, {Jeli{\'c}}, {Reich}, {R{\"o}ttgering}, \&
  {Schnitzeler}}]{vanEck19}
{Van Eck}, C.~L., {Haverkorn}, M., {Alves}, M.~I.~R., {et~al.} 2019, \aap, 623,
  A71

\bibitem[{{van Eck} {et~al.}(2017){van Eck}, {Haverkorn}, {Alves}, {Beck}, {de
  Bruyn}, {En{\ss}lin}, {Farnes}, {Ferri{\`e}re}, {Heald}, {Horellou},
  {Horneffer}, {Iacobelli}, {Jeli{\'c}}, {Mart{\'{\i}}-Vidal}, {Mulcahy},
  {Reich}, {R{\"o}ttgering}, {Scaife}, {Schnitzeler}, {Sobey}, \&
  {Sridhar}}]{vanEck:2017}
{van Eck}, C.~L., {Haverkorn}, M., {Alves}, M.~I.~R., {et~al.} 2017, \aap, 597,
  A98

\bibitem[{{Vansyngel} {et~al.}(2017){Vansyngel}, {Boulanger}, {Ghosh},
  {Wandelt}, {Aumont}, {Bracco}, {Levrier}, {Martin}, \&
  {Montier}}]{Vansyngel17}
{Vansyngel}, F., {Boulanger}, F., {Ghosh}, T., {et~al.} 2017, \aap, 603, A62

\bibitem[{{{\v{S}}nidari{\'c}} {et~al.}(2023){{\v{S}}nidari{\'c}}, {Jeli{\'c}},
  {Mevius}, {Brentjens}, {Erceg}, {Shimwell}, {Piras}, {Horellou}, {Sabater},
  {Best}, {Bracco}, {Ceraj}, {Haverkorn}, {O'Sullivan}, {Turi{\'c}}, \&
  {Vacca}}]{Snidaric23}
{{\v{S}}nidari{\'c}}, I., {Jeli{\'c}}, V., {Mevius}, M., {et~al.} 2023, \aap,
  674, A119

\bibitem[{{Wayth} {et~al.}(2015){Wayth}, {Lenc}, {Bell}, {Callingham},
  {Dwarakanath}, {Franzen}, {For}, {Gaensler}, {Hancock}, {Hindson},
  {Hurley-Walker}, {Jackson}, {Johnston-Hollitt}, {Kapi{\'n}ska}, {McKinley},
  {Morgan}, {Offringa}, {Procopio}, {Staveley-Smith}, {Wu}, {Zheng}, {Trott},
  {Bernardi}, {Bowman}, {Briggs}, {Cappallo}, {Corey}, {Deshpande}, {Emrich},
  {Goeke}, {Greenhill}, {Hazelton}, {Kaplan}, {Kasper}, {Kratzenberg},
  {Lonsdale}, {Lynch}, {McWhirter}, {Mitchell}, {Morales}, {Morgan}, {Oberoi},
  {Ord}, {Prabu}, {Rogers}, {Roshi}, {Shankar}, {Srivani}, {Subrahmanyan},
  {Tingay}, {Waterson}, {Webster}, {Whitney}, {Williams}, \&
  {Williams}}]{Wayth15}
{Wayth}, R.~B., {Lenc}, E., {Bell}, M.~E., {et~al.} 2015, \pasa, 32, e025

\bibitem[{{Wolfire} {et~al.}(1995){Wolfire}, {Hollenbach}, {McKee}, {Tielens},
  \& {Bakes}}]{Wolfire95}
{Wolfire}, M.~G., {Hollenbach}, D., {McKee}, C.~F., {Tielens}, A.~G.~G.~M., \&
  {Bakes}, E.~L.~O. 1995, \apj, 443, 152

\bibitem[{{Wolleben} {et~al.}(2019){Wolleben}, {Landecker}, {Carretti},
  {Dickey}, {Fletcher}, {McClure-Griffiths}, {McConnell}, {Thomson}, {Hill},
  {Gaensler}, {Han}, {Haverkorn}, {Leahy}, {Reich}, \& {Taylor}}]{Wolleben19}
{Wolleben}, M., {Landecker}, T.~L., {Carretti}, E., {et~al.} 2019, \aj, 158, 44

\bibitem[{{Xu} \& {Han}(2019)}]{Xu-Han19}
{Xu}, J. \& {Han}, J.~L. 2019, \mnras, 486, 4275

\bibitem[{{Zaroubi} {et~al.}(2015){Zaroubi}, {Jeli{\'c}}, {de Bruyn},
  {Boulanger}, {Bracco}, {Kooistra}, {Alves}, {Brentjens}, {Ferri{\`e}re},
  {Ghosh}, {Koopmans}, {Levrier}, {Miville-Desch{\^e}nes}, {Montier}, {Pandey},
  \& {Soler}}]{Zaroubi:2015}
{Zaroubi}, S., {Jeli{\'c}}, V., {de Bruyn}, A.~G., {et~al.} 2015, \mnras, 454,
  L46

\bibitem[{{Zucker} {et~al.}(2022){Zucker}, {Goodman}, {Alves}, {Bialy},
  {Foley}, {Speagle}, {Gro{\^I}{\texttwosuperior}schedl}, {Finkbeiner},
  {Burkert}, {Khimey}, \& {Swiggum}}]{Zucker22}
{Zucker}, C., {Goodman}, A.~A., {Alves}, J., {et~al.} 2022, \nat, 601, 334

\end{thebibliography}

\appendix
\section{Magnetic field model } 
\label{App:Bfield}

This appendix details the derivation of the magnetic field model that we use to estimate the Faraday depth from the WNM in Sect.~\ref{sec:LOFAR}. 
The magnetic field model in Eq.~\ref{eq:Bmodel} is written as a sum of the mean magnetic field $\vec{B}_0$ and a random component $\vec{B}_{\rm t}$. We describe how we determine the direction of $\vec{B}_0$ in Sect.~\ref{App_sec:mean_field} and how we compute statistical realizations of $\vec{B}_{\rm t}$ in Sect.~\ref{App_sec:random_field}.
The model presented here was introduced by \citet{PIPXLIV} and \citet{Vansyngel17}. This appendix explains how the specific model we use in this paper was obtained. 

\subsection{Mean magnetic field}
\label{App_sec:mean_field}

\subsubsection{Dust polarization on large angular scales}
\label{App_subsec:model}

The direction of $\vec{B}_0$ is assumed to be uniform. It is defined by the Galactic coordinates $l_0$ and $b_0$ of the unit vector $\hat{B}_0$. 
Introducing the LoS unit vector $\hat{r}$, we derive the component of $\hat{B}_0$ along the LoS, $\hat{B}_{0//}$, and in the plane of the sky, $\hat{B}_{0\perp}$:
\begin{eqnarray}
\begin{aligned}
&\hat{B}_{0//} = \hat{B}_{0}-\hat{B}_{0}\cdot \hat{r} \\
&\hat{B}_{0\perp} = \hat{B}_{0}-\hat{B}_{0//}
\end{aligned}
\end{eqnarray}

To compute the Stokes parameters $I$, $Q$, and $U$, we start
from the integral equations in for instance Sect.~2 of \citet{Regaldo20}. Within the assumption of a uniform $\hat{B}_0$, we get:

\begin{eqnarray}
\begin{aligned}
&Q = \frac{3p_0}{3+2p_0} \, (I+P) \,  \mathrm{cos}^2 \gamma \, \mathrm{cos}\, (2\psi ) \\
&U = -\frac{3p_0}{3+2p_0} \, (I+P) \, \mathrm{cos}^2 \gamma \,\mathrm{sin} \, (2\psi) 
\end{aligned}
\label{eq:Stokes}
\end{eqnarray}
where $P \equiv (Q^2+U^2)^{0.5}$ is the polarized intensity, $p_0$ is a parameter quantifying the intrinsic fraction of dust polarization, $\gamma $ is the angle that $B_0$ makes with the plane
of the sky, and $\psi$ the polarization angle. We note that \citet{PIPXLIV} used a simplified version of these equations without $P$. 
The angles $\gamma$ and $\psi$ are computed from the two following equations:
\begin{eqnarray}
\begin{aligned}
& \mathrm{cos}^2 \, \gamma = 1-(\hat{B}_{0} \cdot \hat{r})^2 \\
& \psi = \pi/2 - \mathrm{acos}\left(\frac{\hat{B}_{0\perp} \cdot \hat{n}}{|\hat{B}_{0\perp}|}\right)
\end{aligned}
\label{eq:angles}
\end{eqnarray}
where $\hat{n}$ is the unit vector perpendicular to $\hat{r}$ within the $\hat{r}$,$\hat{z}$ plane ($\hat{z}$ is the unit vector pointing toward the North Galactic Pole).

\subsubsection{Model fit}
\label{subsec:fit}

Our model of dust polarization on large angular scales defined by Eqs.~\ref{eq:Stokes} and \ref{eq:angles} has three parameters: $l_0$, $b_0$ and $p_0$. We introduce the \Planck\ data and explain how the fit is performed to determine the parameter values with their error bars.

We use the \Planck\ polarization maps at $353\,$GHz produced by the SRoll2 software\footnote{The SRoll2 maps are available \href{http://sroll20.ias.u-psud.fr/sroll20_data.html}{here}.} \citep{Delouis19}. This data release improves the PR3 Legacy polarization maps by correcting systematic effects on large angular scales. We use Healpix all-sky maps at a resolution $ N_\mathrm{side} = 32$ corresponding to a pixel size of $1.8^\circ$ \citep{Gorski05}. 

Eqs.~\ref{eq:Stokes} and \ref{eq:angles} may be combined to compute model maps of the $Q/(I+P)$ and $U/(I+P)$ ratios on a full-sky Healpix grid. This model has three parameters $p_0$, $l_0$ and $b_0$. 

At $ N_\mathrm{side} = 32$, the structure associated with the turbulent component of the magnetic field dominates the data noise and the Cosmic Microwave Background. As this uncertainty scales with (I+P), we compute and fit $Q/(I+P)$ and $U/(I+P)$ maps giving equal weights to all pixels within the unmasked sky area. 

To estimate the error bars of the parameters, we use mock data. The input Stokes maps of dust polarization are those introduced in Appendix~A of \citet{PlanckIII}. We build a set of simulated maps by adding the CMB signal and independent realizations of the data noise from end-to-end simulations of SRoll2 data processing \citep{Delouis19} to these input maps. The dispersion of the parameter values provides the error bars on $l_0$ and $b_0$. The error bar in $p_0$ is dominated by the uncertainty in the offset correction of the total intensity map, which we take to be $40\, \mu$K as in \citet{PlanckXII}. 
We perform the fit over three high Galactic latitude areas in the northern hemisphere L42N, L52N and L62N introduced by \citet{PlanckXI}. The three areas encompass the LOFAR field in Fig.~\ref{fig:LOFAR_map}.
The sky fraction is $f_\mathrm{sky}= 0.24$, 0.28 and 0.33 for LR42N, LR52N and LR62N, respectively.  The results of the fit are listed in Table~\ref{tab:B0_fit}. The model parameters are very close for the three sky regions. To compute the Faraday depths, we use the parameters obtained for the L52N mask.  As dust polarization only constrains the orientation of $\hat{B_0}$, the opposite direction to the values of $l_0$ and $b_0$ in Table~\ref{tab:B0_fit} is an equivalent fit of the \Planck\ data. Among the two possible directions, we choose the one closest to the direction derived from Faraday RMs of nearby pulsars, which show that the mean Galactic magnetic field in the Solar neighborhood points toward longitudes around $l = 90^\circ$ \citep{Xu-Han19}. Dust-polarization data do not constrain the field strength either. Based on the data on pulsars in the northern sky presented by \citet{Sobey19}, we choose the total field strength to be $3\,\mu$G.

\begin{table}[ht!]
{\small
\caption{\Planck\ model fit}       
\label{tab:B0_fit}      
\begin{tabular}{ lcccc}    
\hline\hline  
\\[-1.0ex]
   Sky region & $f_\mathrm{sky}$ & $l_0$ & $b_0$ & $p_0$    \\  
       & (a)  & deg.  & deg. &   \\
    \\[-1.0ex]
   \hline 
   \\[-1.0ex]
LR42N & 0.24 & $62.6 \pm 0.80 $ & $-16.9 \pm 0.82$   & $0.073 \pm^{0.11}_{0.08}$ \\
LR52N & 0.28 & $63.4 \pm 0.76 $  & $-13.5 \pm 0.67 $   & $0.074 \pm^{0.11}_{0.08} $ \\
LR62N & 0.33 & $63.5 \pm 0.70 $   & $-11.3 \pm 0.55 $ & $0.073 \pm^{0.11}_{0.08} $  \\
\\[-1.0ex]
    \hline
\end{tabular} 
\\[1.0ex]}
(a) Sky fraction \\
\end{table}

\subsection{Statistical model of random component }
\label{App_sec:random_field}

To model the random component of the magnetic field $\vec{B}_{\rm t}$, we use the model introduced by \citet{PIPXLIV} and \citet{Vansyngel17}. Then, 
$\vec{B}_{\rm t}$ in the sky direction $\hat{r}$ is computed from
\begin{eqnarray}
\begin{aligned}
&\vec{B}_{\rm t}(\hat{r}) \, = \, \sum_{i=1}^{N} G_i(\hat{r}) \\
&\sqrt{<|\vec{B}_{\rm t}(\hat{r})|^2>} \, = \, f_M \, |B_0|,
\end{aligned}
\label{eq:Bturb}
\end{eqnarray}
where the integration along the LoS is approximated by a sum over a finite number, $N$, of independent Gaussian random fields, $G_i$, generated on the sphere from an angular power spectrum $C_\ell$ scaling as a power-law $\ell ^{-\alpha_M}$ for $\ell \ge 2$.  The mean value and dipole of each $G_i$  are set to zero. The parameter $f_M$ sets the ratio between the standard deviation of |$\vec{B}_{\rm t}$| and |$B_0$|. 

The model has three parameters: $f_M$, $N$, and $\alpha_M$.
We use values: $f_M = 0.9$, $N=7$, and $\alpha_M=2.5$, determined by  \citet{Vansyngel17} from a  fit of \Planck\ power spectra of dust polarization at high Galactic latitudes. These parameter values also provide a good fit of the PDF of the dust polarization fraction $p$, the local dispersion of polarization angles, $\mathcal{S}$, and the anti-correlation between $p$ and $\mathcal{S}$ \citep{PlanckXII}.

\section{Fine structure levels of \NII\ }
\label{App_sec:NII}

This appendix presents an attempt to estimate the EM of ionized gas foreground to the stars using FUSE spectra. Our approach makes use of the collisional excitation of the two fine-structure levels of ionized nitrogen.
As the electron density in the WIM is a few orders of magnitude smaller than the critical densities \citep{Goldsmith15}, the column density of ionized nitrogen $N(\NII^{**}$) at its second excited fine-structure level $3P_2$ is directly related to the gas emission measure.

$\rm N(\NII^{**}$) is estimated from the two absorption lines around 1085\,\AA\ that are within the wavelength range of FUSE spectra. The results are listed in Table~\ref{tab:NII}. 
To derive the emission measure from $\rm N(\NII^{**}$), we used the collisional cross-sections with electrons computed by \citet{Tayal11} for a kinetic temperature of 8000\,K, and the spontaneous radiative decay rates listed by \citet{Goldsmith15}. We use an interstellar nitrogen abundance of $\rm log(N/H)_{ISM}=-4.2$. Then we find that EM$_{\NII^{**}}$ (pc cm$^{-6}) = \rm N(\NII^{**}) / 1.8 \,10^{12} cm^{-2}$.

For comparison, we also list in Table~\ref{tab:NII} the H$_\alpha$ line intensities $I(\mathrm{H}\alpha$) at the position of the stars that we measured using the Wisconsin H-alpha Mapper (WHAM) sky survey \citep{Haffner03}. The total emission measure in the direction of the stars was derived from $I(\mathrm{H}\alpha)$ for an electron temperature of 8000\,K. 

For most LoSs, we derive upper limits on $\rm N(\NII^{**})$ of the order of $ 10^{13}\,\mathrm{cm}^{-2}$, which translates into upper limits for EM$_{\NII^{**}}$ that are
significantly larger than the total emission measure derived from $I(\mathrm{H}\alpha)$. 
For the two cases, HD113001 and Ton 102, which yielded actual detections of $\rm N(\NII^{**})$ absorption, the inferred EM values far exceeded those determined from the H$\alpha$ emission. It is possible that these large contributions of $\rm N(\NII^{**})$ arise from the \HII\ regions surrounding the stars that subtend an angle much smaller than the $1^\circ$ beam width of the WHAM survey.

\begin{table}

%\centering
{\small
    \caption{Column densities of \NII$^{**}$ and emission measures}
\label{tab:NII}
   \begin{tabular}{lcccc}
    \hline
    \hline
    \\[-1.0ex]
 Star    & log($N(\NII^{**})$) & EM$_{\NII}$ & $I(\mathrm{H}\alpha$)  & EM$_{\mathrm{H}\alpha}$ \\
   & $\mathrm{cm}^{-2}$ & $\mathrm{pc\, cm}^{-6}$ & R &$\mathrm{pc\, cm}^{-6}$ \\
   & (a) & (b) & (c) & (d)\\
    \\[-1.0ex]
    \hline \\[-1.0ex]
HZ43A & $< 13.2$ & $< 8.8$ & $0.20\pm 0.03$ & $0.4 \pm 0.1$ \\
PG1544+488 & $< 13.2$& $< 8.8$& $0.59\pm 0.04$ & $1.3 \pm 0.1$ \\
PG1610+519 & $< 13.3$&$< 11.0$ & $0.63\pm 0.03$ & $1.4 \pm 0.1$ \\ 
HD113001 &$14.55\pm^{0.3}_{0.25}$ & $200\pm^{190}_{90}$ & $0.53\pm 0.03$ & $1.2 \pm 0.1$\\
Ton102 & $14.8^{+0.9}_{-0.4}$&$ 350\pm^{2400}_{110}$& $0.63\pm 0.03$ & $ 1.4 \pm 0.1$\\
PG1051+501 & $< 14.0$& $< 56$& $-0.08 \pm 0.04$ & $< 0.2$\\
PG0952+519 & $$< 13.0& $$< 5.6& $0.83\pm 0.05$ & $1.9 \pm 0.1$\\
Feige34 & $< 13.0$ & $< 5.6$ & $0.94\pm 0.04$ & $2.1 \pm 0.1$\\
HD93521 & $< 12.7$ & $< 2.8$ & $0.49\pm 0.04$ & $1.1 \pm 0.1$ \\
     \\[-1.0ex]
    \hline
\end{tabular} 
\\[1.0ex]}
a:  Column density of \NII$^{**}$.\\
b : Emission measure derived from the column density of \NII$^{**}$, assuming T $= 8000 K$   \\
c: Intensity of the H$\alpha$ line measured with the WHAM data in units of Rayleigh.\\
d: Total emission measure derived from the H$\alpha$ line intensity. \\
\end{table}

\end{document}